\documentclass[twocolumn,times]{aastex61}
\usepackage{amsmath}

\shorttitle{Microflare heating observed by {\it NuSTAR} \& {\it Hinode}/XRT}
\shortauthors{Wright et al.}

\begin{document}

\title{Microflare Heating of a Solar Active Region Observed with {\it N\lowercase{u}STAR}, {\it Hinode}/XRT, and {\it SDO}/AIA}
\author[0000-0001-9021-611X]{Paul J. Wright}
\affiliation{SUPA School of Physics \& Astronomy, University of Glasgow, Glasgow G12 8QQ, UK}

\author{Iain G. Hannah}
\affiliation{SUPA School of Physics \& Astronomy, University of Glasgow, Glasgow G12 8QQ, UK}

\author{Brian W. Grefenstette} 
\affiliation{Cahill Center for Astrophysics, 1216 E. California Blvd, California Institute of Technology, Pasadena, CA 91125, USA}

\author{Lindsay Glesener}
\affiliation{School of Physics \& Astronomy, University of Minnesota - Twin Cities, Minneapolis, MN 55455, USA}

\author{S{\"a}m Krucker}
\affiliation{Space Sciences Laboratory University of California, Berkeley, CA 94720, USA}
\affiliation{University of Applied Sciences and Arts Northwestern Switzerland, 5210 Windisch, Switzerland}
 
\author{Hugh S. Hudson}
\affiliation{Space Sciences Laboratory University of California, Berkeley, CA 94720, USA}
\affiliation{SUPA School of Physics \& Astronomy, University of Glasgow, Glasgow G12 8QQ, UK}

\author{David M. Smith}
\affiliation{Santa Cruz Institute of Particle Physics and Department of Physics, University of California, Santa Cruz, CA 95064, USA}

\author{Andrew J. Marsh} 
\affiliation{Santa Cruz Institute of Particle Physics and Department of Physics, University of California, Santa Cruz, CA 95064, USA}

\author{Stephen M. White}
\affiliation{Air Force Research Laboratory, Space Vehicles Directorate, 3550 Aberdeen Ave SE, Kirtland AFB, NM 87117, USA}

\author{Matej Kuhar}
\affiliation{University of Applied Sciences and Arts Northwestern Switzerland, 5210 Windisch, Switzerland}

\correspondingauthor{Paul J. Wright}
\email{paul.wright@glasgow.ac.uk}

\begin{abstract}

{\it NuSTAR} is a highly sensitive focusing hard X-ray (HXR) telescope and has observed several small microflares in its initial solar pointings. In this paper, we present the first joint observation of a microflare with {\it NuSTAR} and {\it Hinode}/XRT on 2015 April 29 at $\sim$11:29 UT. This microflare shows heating of material to several million Kelvin, observed in Soft X-rays (SXRs) with {\it Hinode}/XRT, and was faintly visible in Extreme Ultraviolet (EUV) with {\it SDO}/AIA. For three of the four {\it NuSTAR} observations of this region (pre-, decay, and post phases) the spectrum is well fitted by a single thermal model of $3.2-3.5$ MK, but the spectrum during the impulsive phase shows additional emission up to $10$ MK, emission equivalent to A0.1 {\it GOES} class. We recover the differential emission measure (DEM) using {\it SDO}/AIA, {\it Hinode}/XRT, and {\it NuSTAR}, giving unprecedented coverage in temperature. We find the pre-flare DEM peaks at $\sim3$ MK and falls off sharply by $5$ MK; but during the microflare's impulsive phase the emission above $3$ MK is brighter and extends to $10$ MK, giving a heating rate of about $2.5 \times 10^{25}$ erg s$^{-1}$. As the {\it NuSTAR} spectrum is purely thermal we determined upper-limits on the possible non-thermal bremsstrahlung emission. We find that for the accelerated electrons to be the source of the heating requires a power-law spectrum of $\delta \ge 7$ with a low energy cut-off $E_{c} \lesssim 7$ keV. In summary, this first {\it NuSTAR} microflare strongly resembles much more powerful flares.

\end{abstract}

\keywords{Sun: X-rays, gamma rays --- Sun: activity --- Sun: corona}

\section{Introduction}

Solar flares are rapid releases of energy in the corona and are typically characterised by impulsive emission in Hard X-rays (HXRs) followed by brightening in Soft X-rays (SXRs) and Extreme Ultraviolet (EUV) indicating that electrons have been accelerated as well as material heated.

Flares are observed to occur over many orders of magnitude, from large X-Class {\it GOES} ({\it Geostationary Operational Environmental Satellite}) flares down to A-class microflares. Observations from {\it RHESSI} \citep[{\it Reuven Ramaty High Energy Solar Spectroscopic Imager};][]{2002SoPh..210....3L} have shown that microflares occur exclusively in active regions (ARs), like larger flares, as well as heating material $> 10$ MK and accelerating electrons to $> 10$ keV \citep{2008ApJ...677.1385C, 2008ApJ...677..704H, 2011SSRv..159..263H}. Although energetically these events are about six orders of magnitude smaller than large flares it shows that the same physical processes are at work to impulsively release energy. There should be smaller events beyond {\it RHESSI}'s sensitivity but so far there have only been limited SXR observations from {\it SphinX} \citep{2011SoSyR..45..189G} or indirect evidence of non-thermal emission from {\it IRIS} obser vations \citep[e.g.][]{2014Sci...346B.315T}. There are also energetically smaller events observed in thermal EUV/SXR emission that occur outside ARs \citep{1997ApJ...488..499K,2000ApJ...529..554P, 2000ApJ...535.1047A}.

Smaller flares occur considerably more often than large flares with their frequency distribution behaving as a negative power-law \citep[e.g.][]{2011SSRv..159..263H}. It is not clear how small flare-like events can be, with \citet{1988ApJ...330..474P} suggesting that small scale reconnection events (``nanoflares'') are on the order of $\sim$10$^{24}$ erg. However at this scale flares are likely too small to be individually observed, and only the properties of the unresolved ensemble could be determined \citep{1975ApJ...199L..53G}. Nor it is clear if the flare frequency distribution is steep enough \citep[requiring $\alpha>$2,][]{1991SoPh..133..357H} so that there are enough small events to keep the solar atmosphere consistently heated. It is therefore crucial to probe how small flares can be while still remaining distinct, and how their properties relate to flares and microflares.

With the launch of the {\it Nuclear Spectroscopic Telescope ARray} \citep[{{\it NuSTAR}};][]{2013ApJ...770..103H}, HXR ($2.5 - 78$ keV) observations of faint, previously undetectable solar sources can be obtained. In comparison to {\it RHESSI}, {\it NuSTAR} has over $10\times$ larger effective area and a much smaller background counting rate. However {\it NuSTAR} was designed for astrophysical observations and is therefore not optimised for observations of the Sun. This leads to various technical challenges \citep[see][]{2016BWG}, 
but {\it NuSTAR} is nevertheless a unique instrument for solar observations and has pointed at the Sun several times. {\it NuSTAR} has observed several faint sources from quiescent ARs \citep{2016ApJ...820L..14H} and emission from an occulted flare, in the EUV late-phase \citep{2017ApJ...835....6K}. {\it NuSTAR} has also observed several small microflares during its solar observations, one showing the time evolution and spectral emission \citep{2017Glesener}.

In this paper we present {\it NuSTAR} imaging spectroscopy of the first microflare jointly observed with {\it Hinode}/XRT \citep{2007SoPh..243....3K, 2007SoPh..243...63G} and {\it SDO}/AIA \citep{2012SoPh..275....3P,2012SoPh..275...17L}. This microflare occurred on 2015 April 29 within AR12333, and showed distinctive loop heating visible with {\it NuSTAR}, {\it Hinode}/XRT, and the hottest EUV channels of {\it SDO}/AIA up to $10$ MK. We first present an overview of {\it SDO}/AIA, and {\it Hinode}/XRT observations in \S\ref{overview}, followed by {\it NuSTAR} data analysis in \S\ref{sec:NuSTAR}. In \S\ref{sec:multit} we concentrate on the impulsive phase of the microflare and perform differential emission measure analysis. Finally in \S\ref{sec:meng} we look at the microflare energetics in terms of thermal, and non-thermal emission.
\section{{\it SDO}/AIA and {\it Hinode}/XRT Event Overview\label{overview}}
\begin{figure*}
\centering
	\includegraphics[width=2.0\columnwidth]{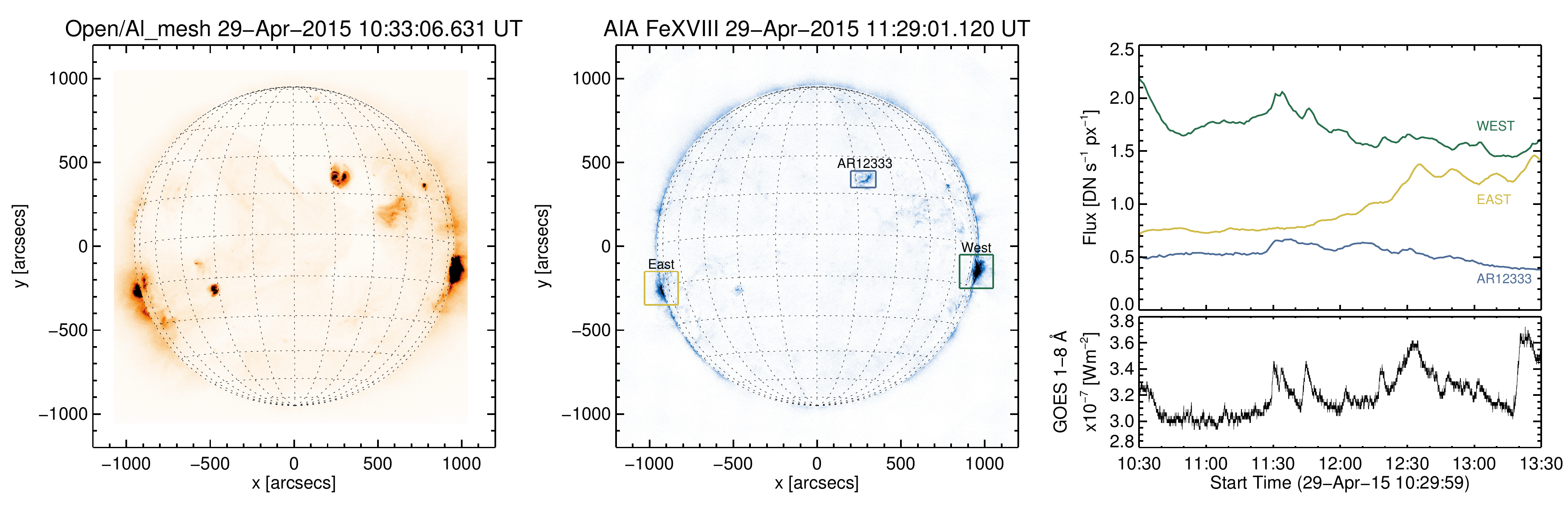} 
	\caption{\label{fig:over}Overview of the {\it SDO}/AIA 94\AA~\ion{Fe}{18} conditions during the times of the {\it NuSTAR} observations, and {\it Hinode}/XRT prior to the AR12333 microflare onset. (Left) full-disk image from {\it Hinode}/XRT one hour prior to the microflare onset. (Middle) full-disk {\it SDO}/AIA 94\AA~\ion{Fe}{18} image at the peak of the microflare impulsive phase with the ARs indicated. The {\it SDO}/AIA 94\AA~\ion{Fe}{18} light curves from these three regions are shown in comparison to the full-disk {\it GOES} 1-8\AA~SXR flux (right). All of the regions are producing several microflares during these times, but those from AR12333 are hidden in the {\it GOES} lightcurve as those from the two limb regions are brighter.
}
\end{figure*}

The microflare from AR12333 occurred during a time when there were two brighter ARs on the disk, as can be seen in Figure~\ref{fig:over}. Both of these ARs, on either limb, were producing microflares that dominate the overall {\it GOES} 1-8\AA~SXR light curve (Figure~\ref{fig:over}, right panels). {\it GOES} is spatially integrated, but the contributions from each region can be determined by using the hotter \ion{Fe}{18} component of {\it SDO}/AIA 94\AA~images. The \ion{Fe}{18} line contribution to the {\it SDO}/AIA 94\AA~channel peaks at $log_{10}T = 6.85$ K ($\sim7$ MK), and can be recovered using a combination of the {\it SDO}/AIA channels \citep{2011ApJ...736L..16R, 2012ApJ...759..141W, 2012ApJ...750L..10T, 2013A&A...558A..73D}. Here we use the approach of \citet{2013A&A...558A..73D}

\begin{equation}
\begin{split}
F(\text{\ion{Fe}{18}}) \approx&~F(\text{94\AA}) \\ 
&- \frac{F(\text{211\AA})}{120.} - \frac{F(\text{171\AA})}{450.}, 
\label{eqn:Fe18}
\end{split}
\end{equation}

\noindent where $F(\text{\ion{Fe}{18}})$ is the \ion{Fe}{18} flux [DN s$^{-1}$ px$^{-1}$] and $F(\text{94\AA})$, $F(\text{171\AA})$, $F(\text{211\AA})$, are the equivalent fluxes in the {\it SDO}/AIA 94\AA, 171\AA, and 211\AA~channels.

\begin{figure*}
\centering
	\includegraphics[width=2.0\columnwidth]{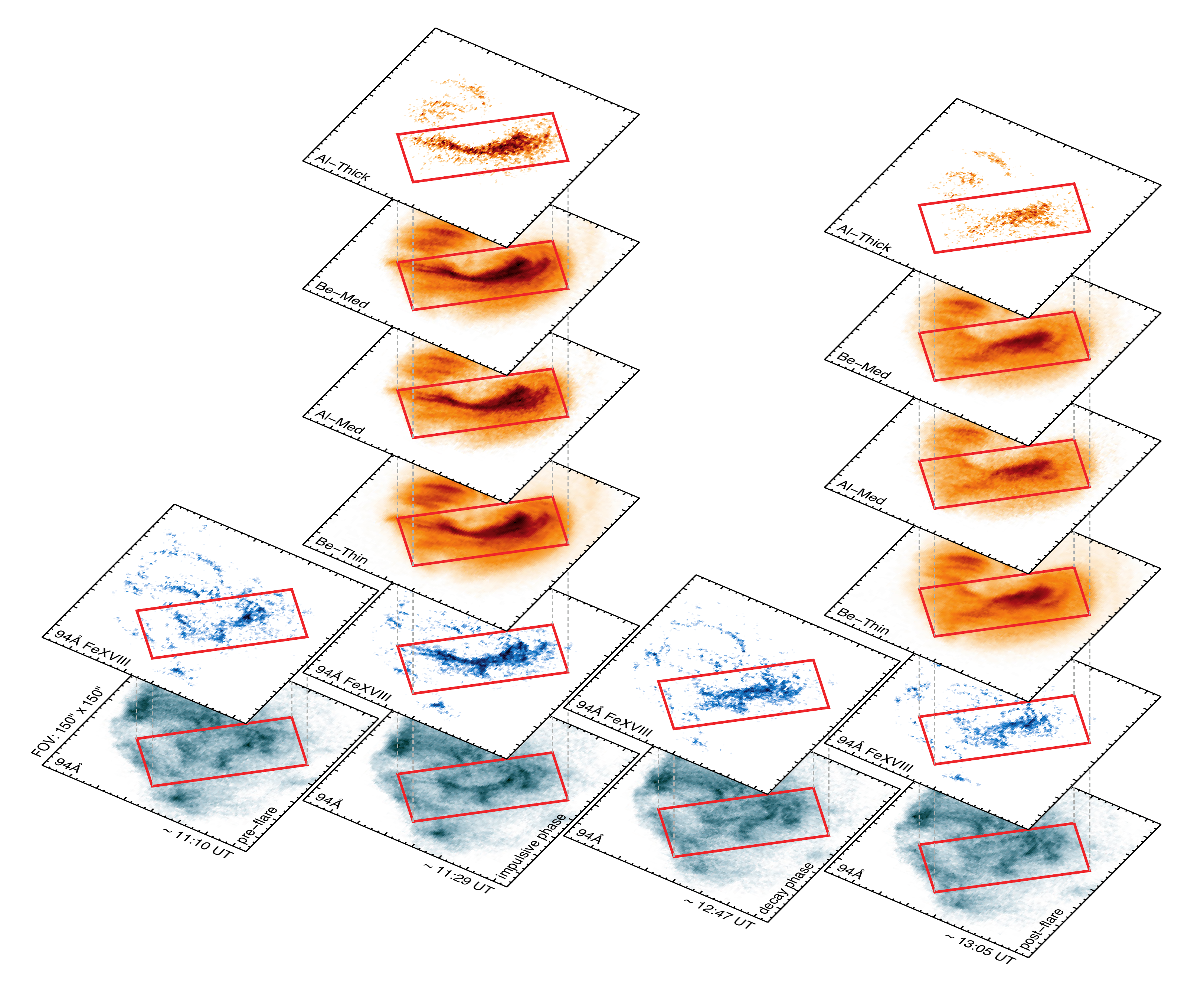}\\
	\caption{\label{fig:aiaxrtmap}Comparison of AR12333 from {\it SDO}/AIA and {\it Hinode}/XRT at the times of {\it NuSTAR} observations (pre-flare, $\sim$ 11:10 UT; impulsive phase, $\sim$ 11:29 UT; decay phase, $\sim$ 12:47 UT; and post-flare, $\sim$ 13:05 UT). The loop region ($95\arcsec \times 45\arcsec$) used for the light curves and DEM analysis is over plotted as a red rectangle. The loop region is faintly observable in {\it SDO}/AIA 94\AA~with the structure well-recovered in the {\it SDO}/AIA 94\AA~\ion{Fe}{18} and SXR channels.}
\end{figure*}

\begin{figure}
\centering
	\includegraphics[width=0.9\columnwidth]{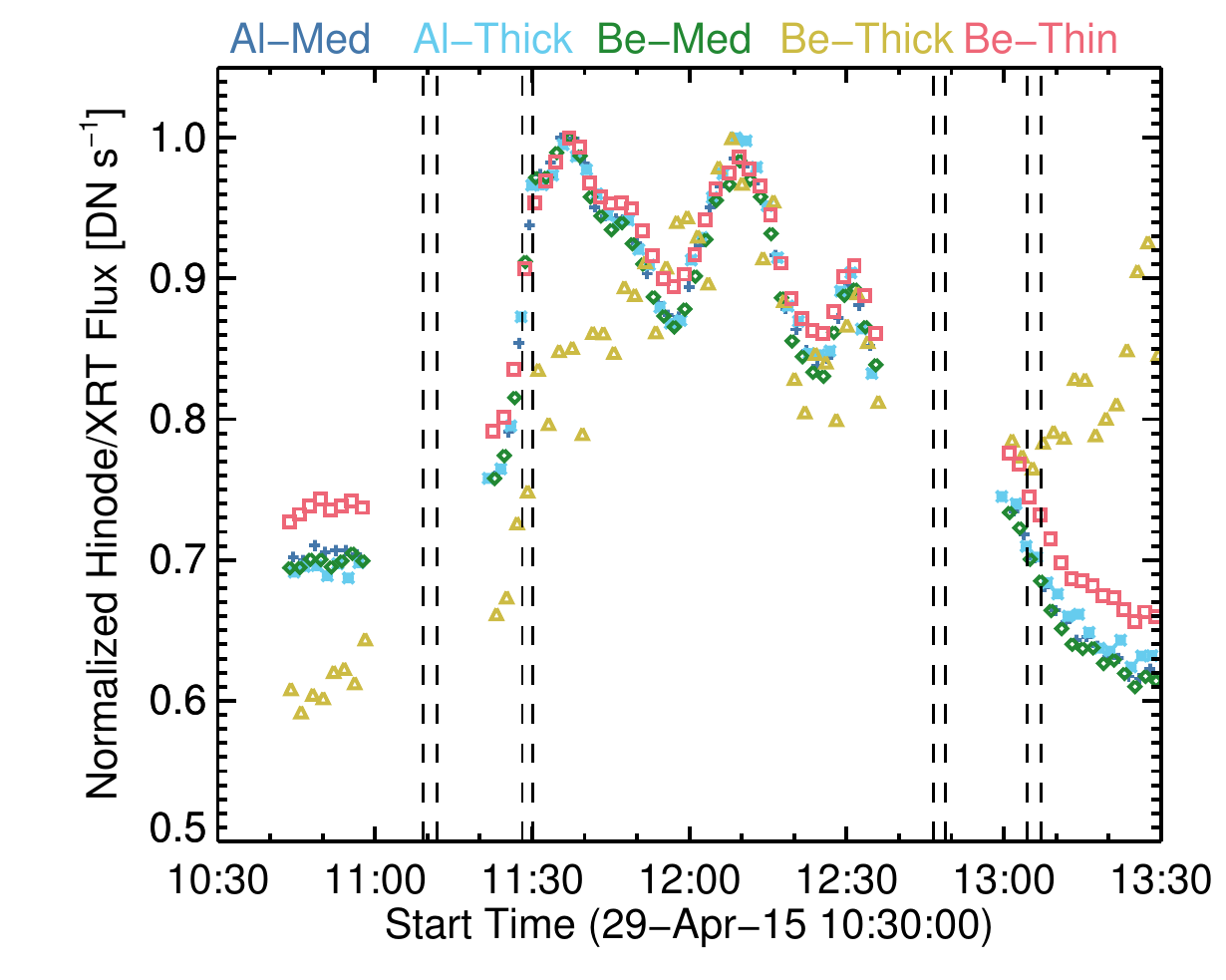} \\
	\includegraphics[width=0.9\columnwidth]{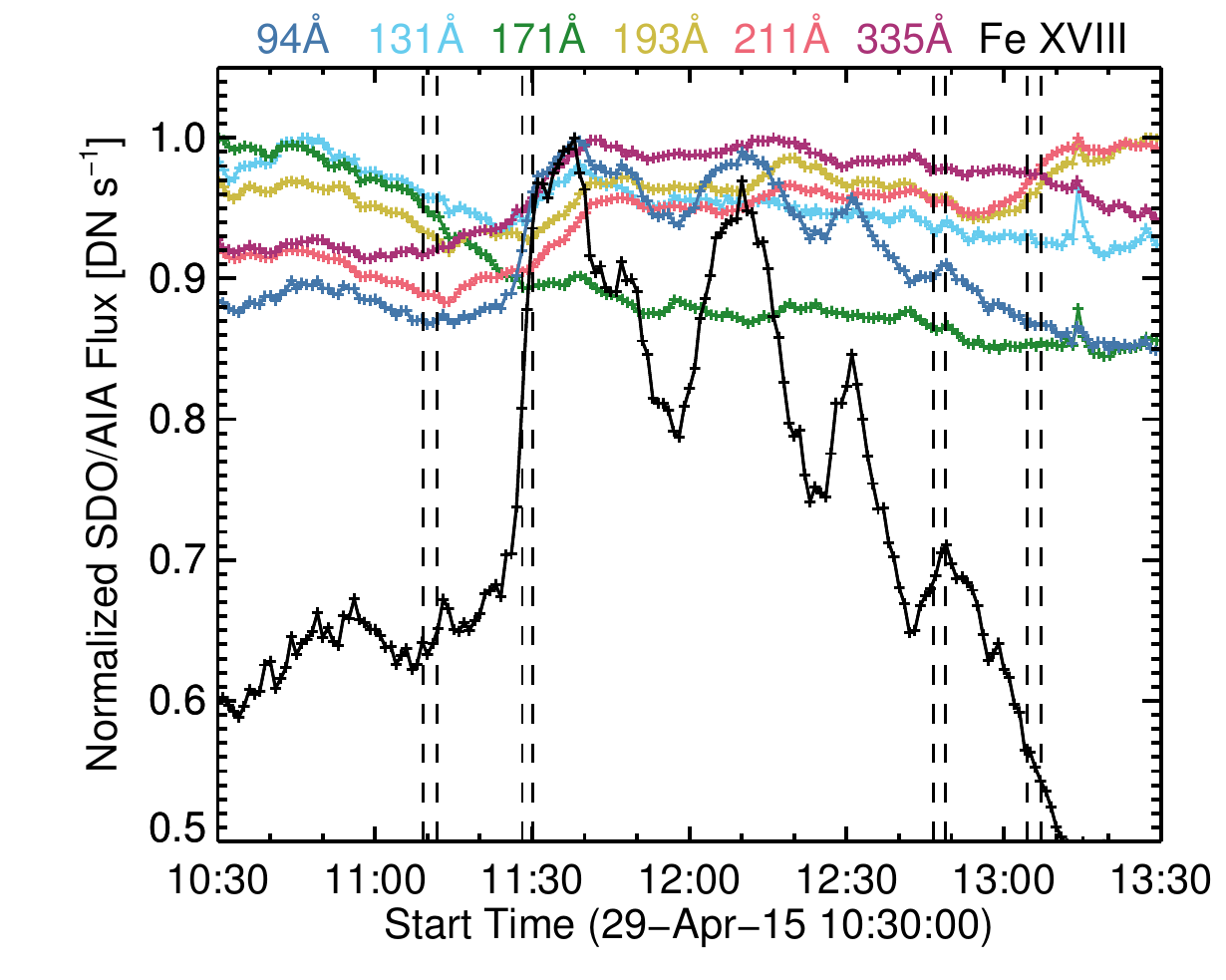}
	 \caption{\label{fig:timeprofile}Time profiles of the different {\it Hinode}/XRT (top) and {\it SDO}/AIA (bottom) channels from the loop region of AR12333 shown in Figure \ref{fig:aiaxrtmap}. The vertical bars indicate the four time periods of the {\it NuSTAR} observation of the same region. The gaps in the {\it Hinode}/XRT light curves are due to incomplete coverage.}
\end{figure}

{\it Hinode}/XRT observed AR12333 in a high cadence mode ($\sim2-3$ minutes), cycling through five different filter channels centered on this region. Full-disk synoptic images were obtained before and after this observation mode (Figure~\ref{fig:over}). Figure~\ref{fig:aiaxrtmap} shows the main loops of the region rapidly brightening, indicating that energy is being released to heat these loops. This is apparent in the SXR channels from {\it Hinode}/XRT and {\it SDO}/AIA 94\AA~\ion{Fe}{18}, but not the cooler EUV channels from {\it SDO}/AIA, so we conclude that the heating is mostly above $3$ MK. For the $95\arcsec \times 45\arcsec$ loop region shown in Figure~\ref{fig:aiaxrtmap} we produce the time profile of the microflares in each of these SXR and EUV channels, shown in Figure~\ref{fig:timeprofile}. These light curves have been obtained after processing via the instrument preparation routines, de-rotation of the solar disk (to $\sim$11:29 UT), and manual alignment of {\it Hinode}/XRT Be-Thin to the $1\arcsec$ down-sampled {\it SDO}/AIA 94\AA~\ion{Fe}{18} data. Here we again see that the microflare activity is only occurring in the channels sensitive to the hottest material, i.e. the SXR ones from {\it Hinode}/XRT and {\it SDO}/AIA 94\AA~\ion{Fe}{18}. This activity is in the form of three distinctive peaks with the first, and largest, impulsively starting at $\sim$ 11:29 UT. This is clear in the SXR (with the exception of the low signal-to-noise {\it Hinode}/XRT Be-Thick channel), and {\it SDO}/AIA 94\AA~\ion{Fe}{18} lightcurves, all showing similar time profiles.

\section{{\it N\lowercase{u}STAR} Data Analysis\label{sec:NuSTAR}}
{\it NuSTAR} is an imaging spectrometer with high sensitivity to X-rays over $2.5$ to $78$ keV \citep{2013ApJ...770..103H}. {\it NuSTAR} consists of two identical telescopes each with the same $12\arcmin \times 12\arcmin$ field of view \citep{2015ApJS..220....8M} and is composed of Wolter-I type optics that directly focus X-rays onto the focal-plane modules (FPMA and FPMB) $10$ m behind. These focal-plane modules each comprise of CdZnTe detectors with $64 \times 64$ pixels providing the time, energy, and location of the incoming X-rays. The readout time per event is 2.5 ms, and {\it NuSTAR} accepts a maximum throughput of $400$ counts s$^{-1}$ for each focal-plane module. This makes {\it NuSTAR} highly capable of observing weak thermal or non-thermal X-ray sources from the Sun \citep{2016BWG}. However, as it is optimised for astrophysics targets solar pointings have limitations. In particular, the low detector readout and large effective area produce high detector deadtime even for modest levels of solar activity, restricting the spectral dynamic range, only detecting X-rays at the lowest energies \citep{2016ApJ...820L..14H,2016BWG}. {\it NuSTAR} solar observations are therefore from times of weak solar activity, ideally when the {\it GOES} $1-8$\AA~ flux is below B-level. An overview of the initial {\it NuSTAR} solar pointings, that began in late 2014, and details of these restrictions is available in \citet{2016BWG}. An up-to-date quicklook summary is also available online\footnote{\url{http://ianan.github.io/nsigh_all/}}.

\begin{figure*}
\centering
	\includegraphics[width=0.45\columnwidth]{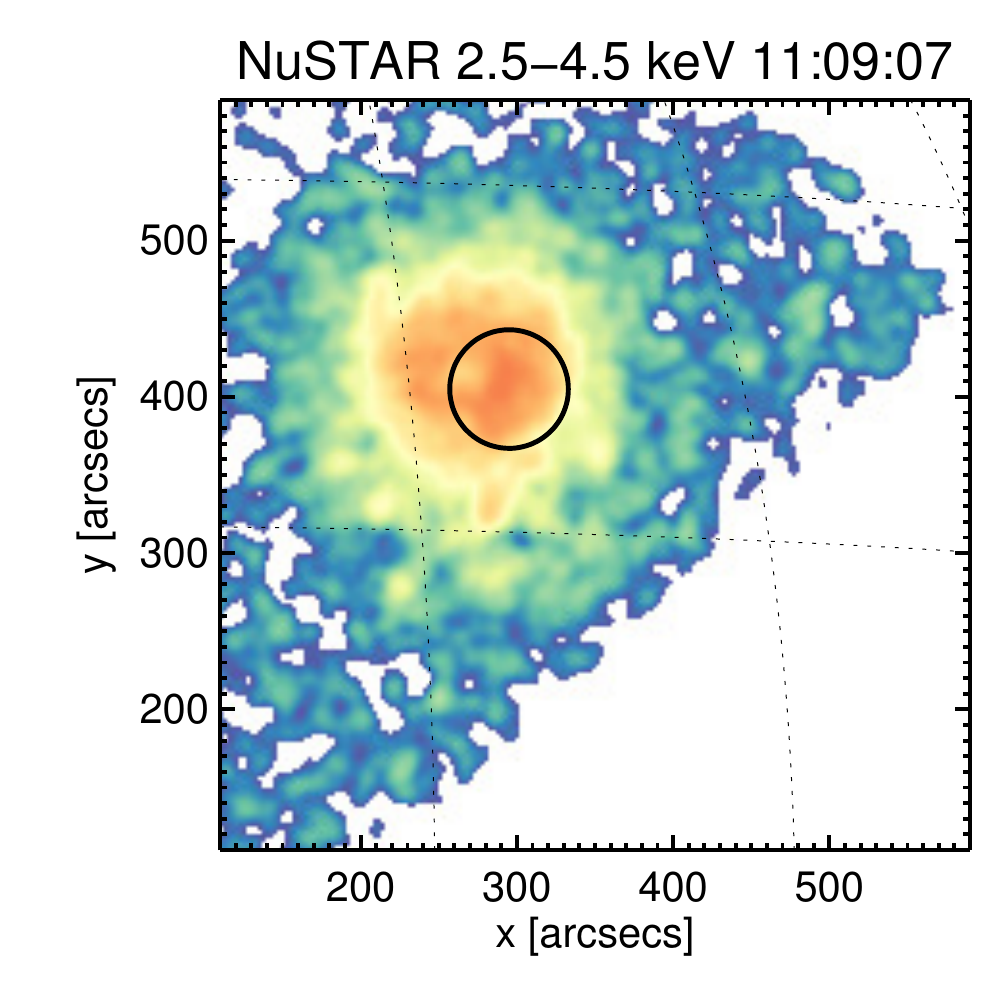}
	\includegraphics[width=0.45\columnwidth]{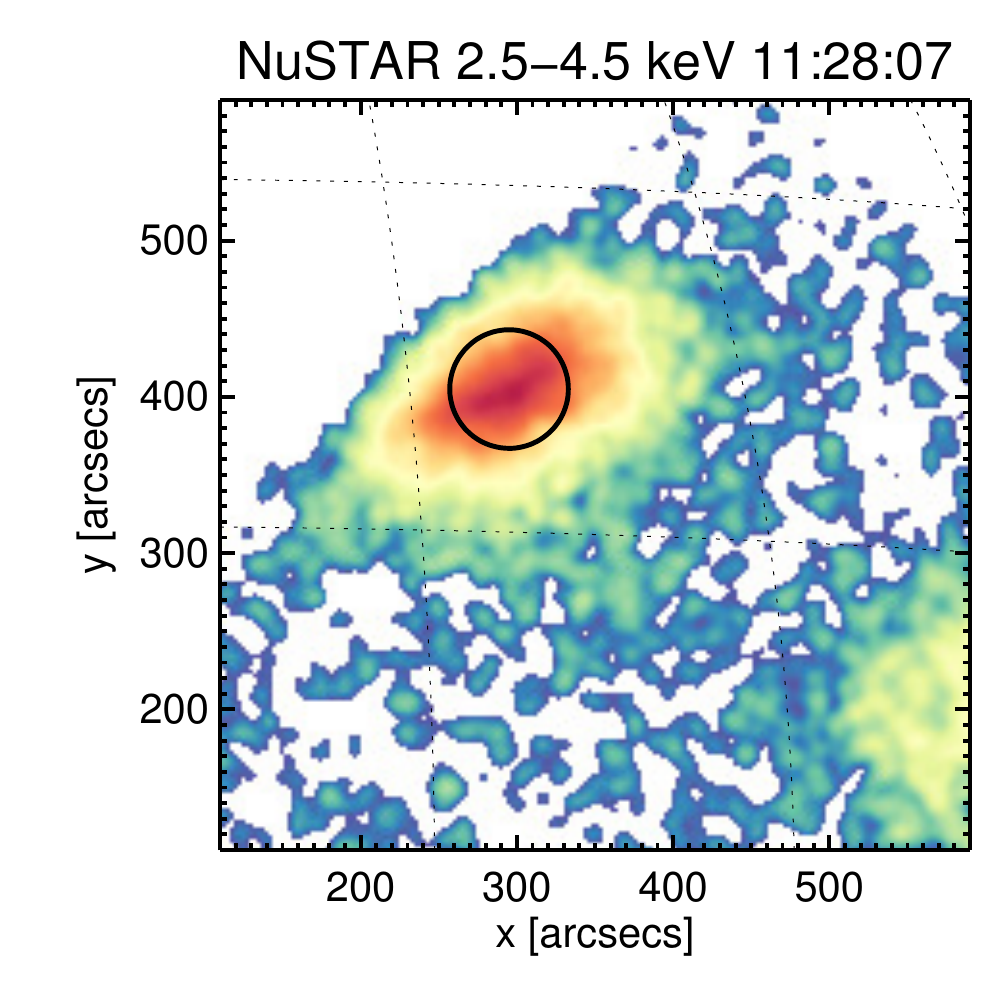}
	\includegraphics[width=0.45\columnwidth]{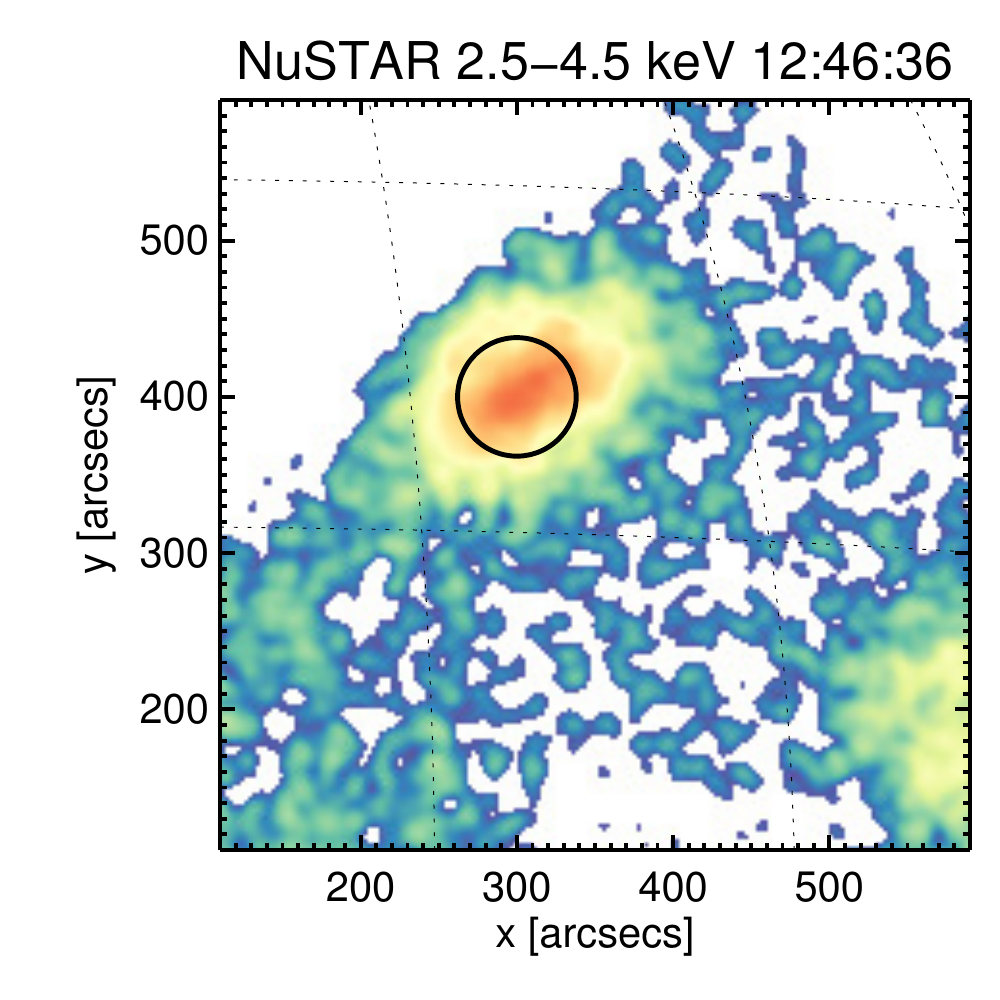}
	\includegraphics[width=0.45\columnwidth]{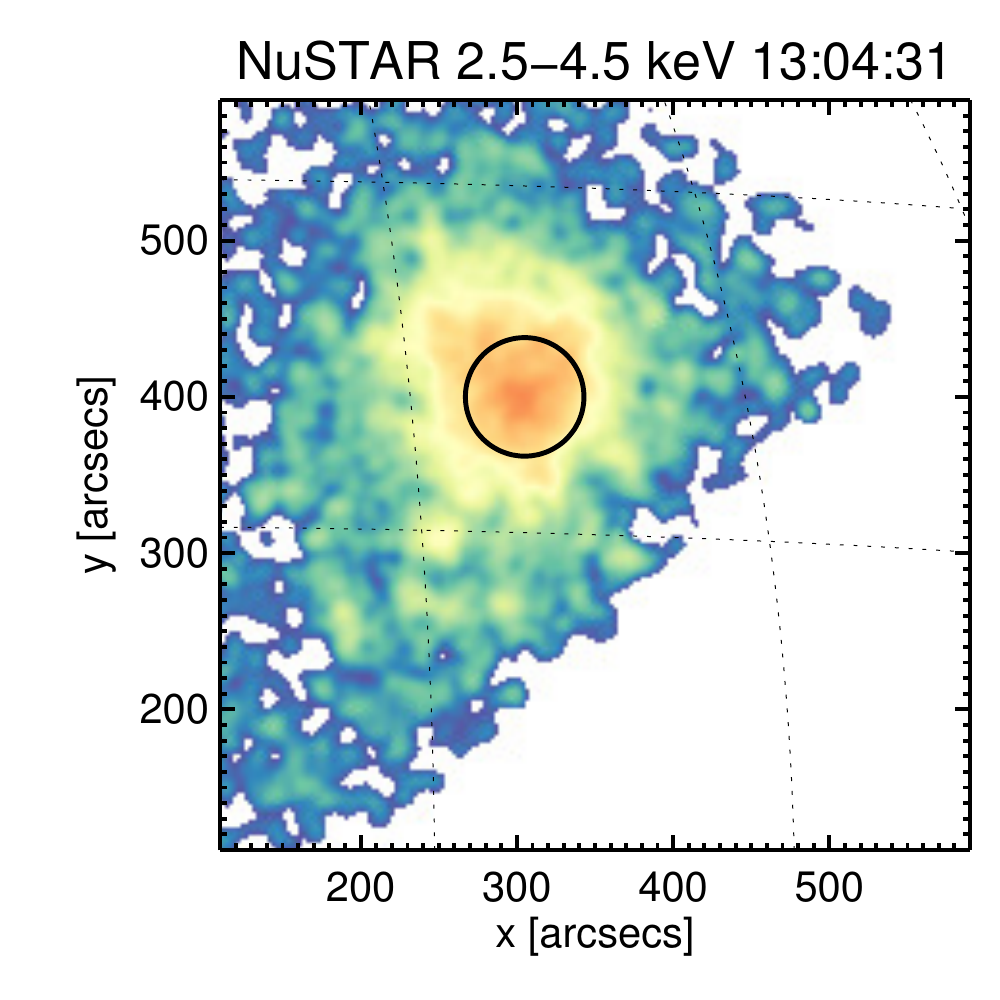}
	\caption{\label{fig:nsmap}{\it NuSTAR} $2.5-4.5$ keV maps for the four time intervals it observed AR12333. These maps have been shifted to match the position of the {\it SDO}/AIA 94\AA~\ion{Fe}{18} maps, shown in Figure~\ref{fig:ns_shift}. The black circles indicate the regions chosen for spectral fitting, shown in Figure~\ref{fig:ns_xspec1}. Note that the same colour scaling is used in all these maps.}
\end{figure*}

The observations reported here are based around the fourth {\it NuSTAR} solar pointing, consisting of two orbits of observations covering 2015 April 29 10:50 to 11:50 and 12:27 to 13:27 \citep{2016BWG}. {\it NuSTAR} completed a full disk mosaic observation in each orbit consisting of 17 different pointings: the field of view requires 16 different pointings to cover the whole Sun, with some overlaps between each mosaic tile, followed by an additional disk centre pointing \citep[see Figure 4][]{2016BWG}. This resulted in {\it NuSTAR} observing AR12333 four times, each lasting for a few minutes. These times are shown in Figure~\ref{fig:timeprofile}. These data were processed using the {\it NuSTAR} Data Analysis software v1.6.0 and {\it NuSTAR} CALDB 20160502\footnote{\url{http://heasarc.gsfc.nasa.gov/docs/NuSTAR/analysis/}}, which produces an event list for each pointing. We use only single-pixel (``Grade 0'') events \citep{2016BWG}, to minimize the effects of pile-up. Figure~\ref{fig:nsmap} shows the resulting {\it NuSTAR} $2.5-4.5$ keV image for each of the four pointings and these images are a combination of both FPMA and FPMB with $\sim$7$\arcsec$ Gaussian smoothing as the pixel size is less than the full width at half maximum (FWHM) of the optics.

\begin{figure*}
\centering
	\includegraphics[width=0.45\columnwidth]{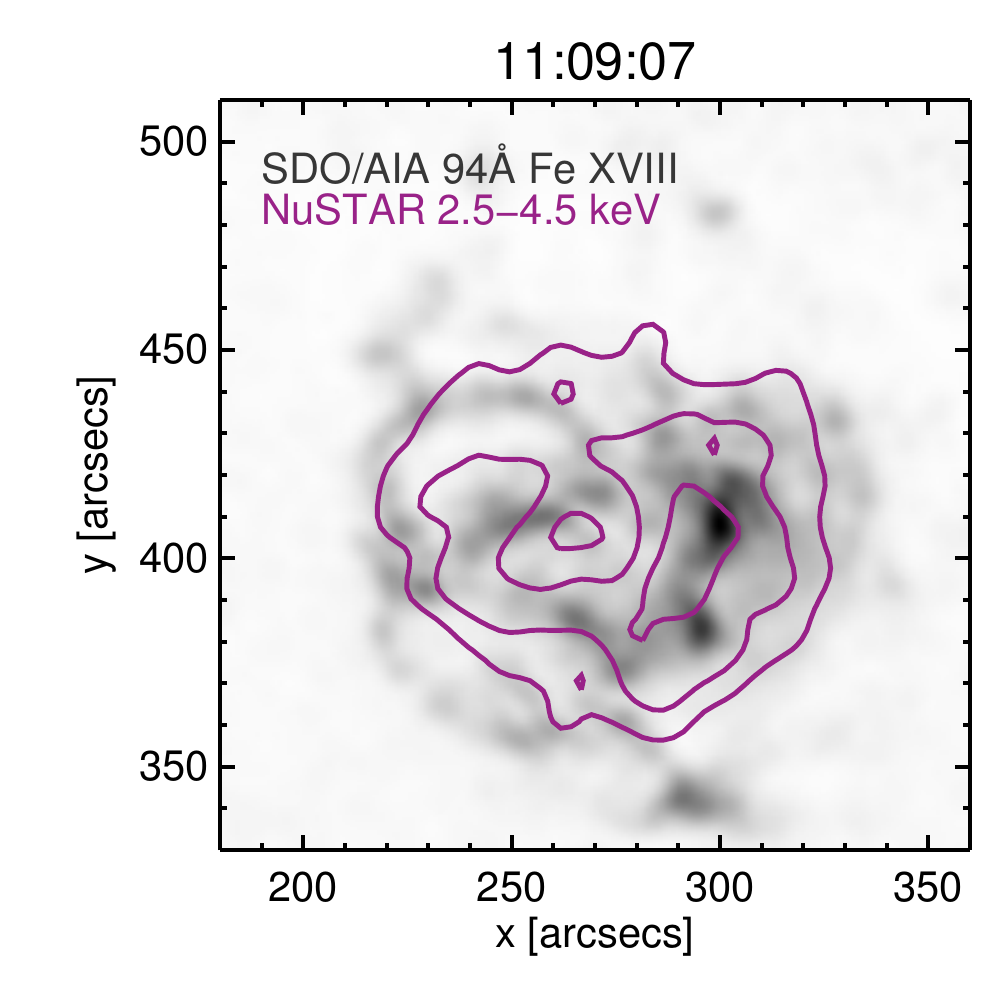}
	\includegraphics[width=0.45\columnwidth]{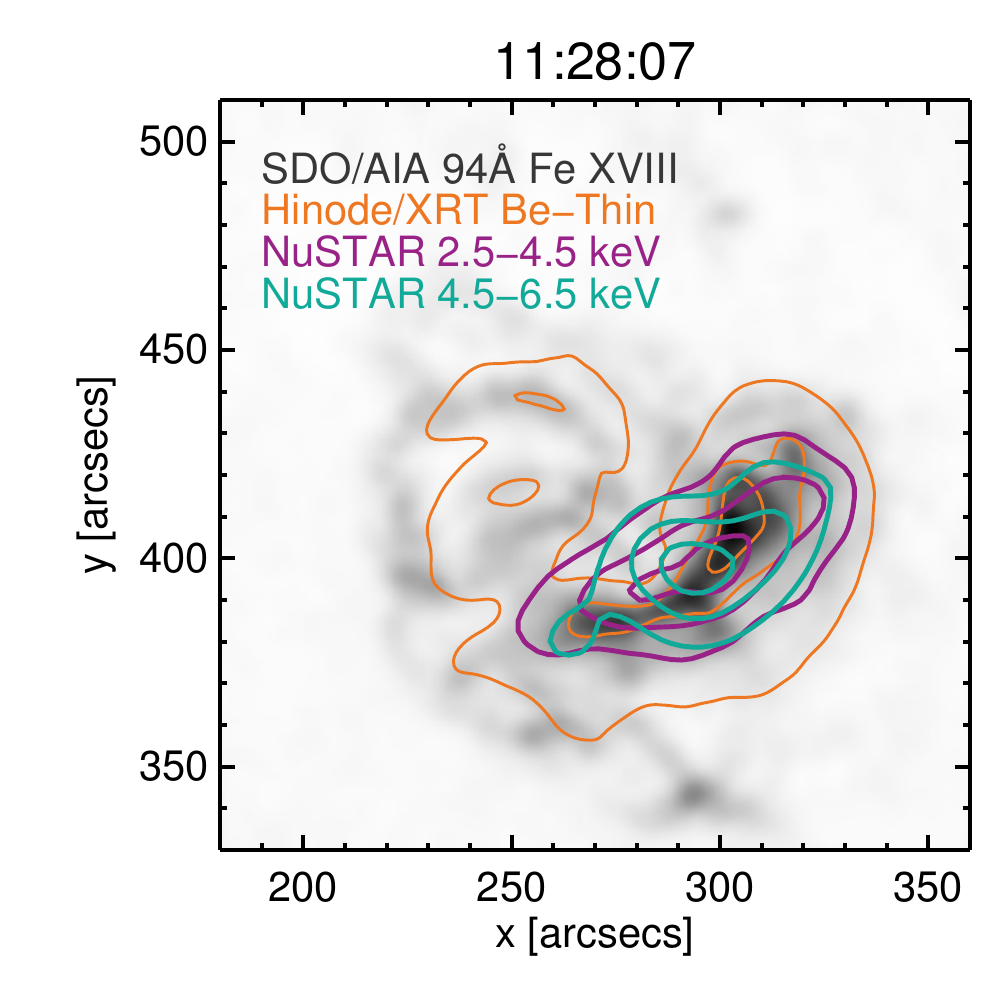}
	\includegraphics[width=0.45\columnwidth]{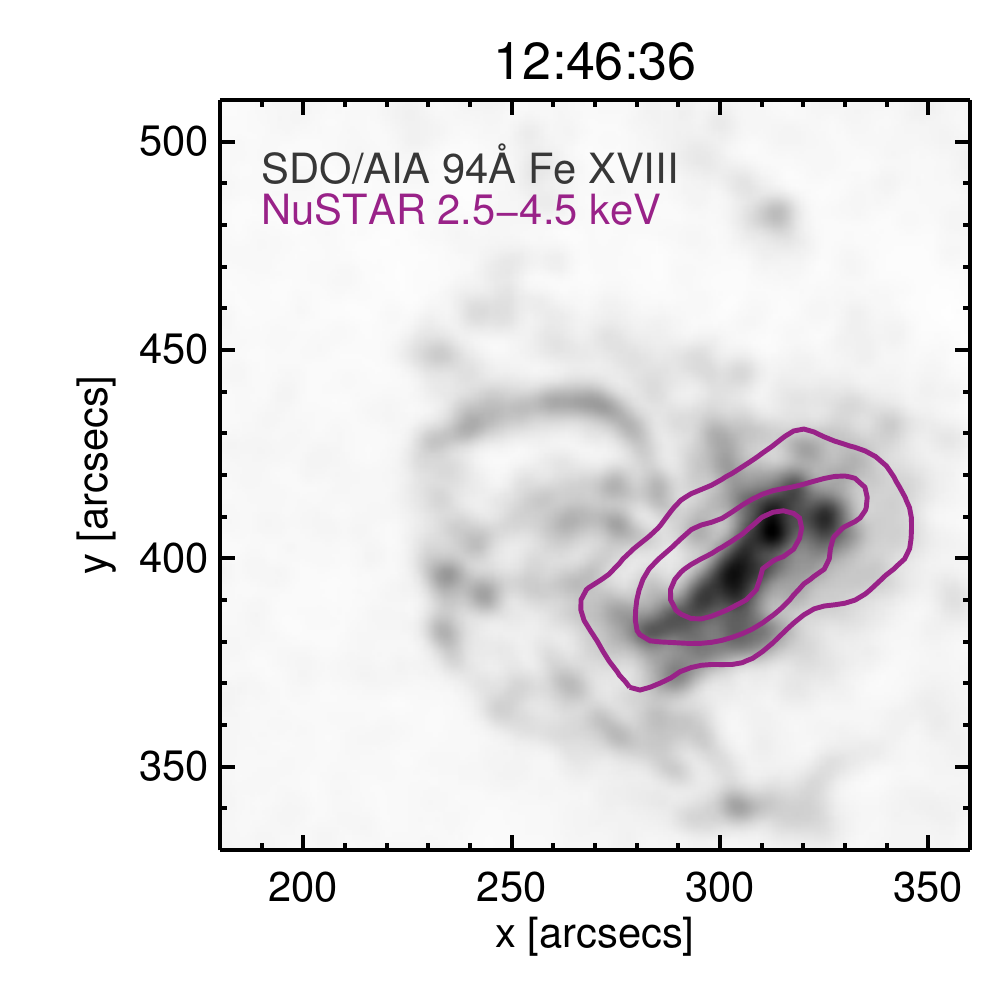}
	\includegraphics[width=0.45\columnwidth]{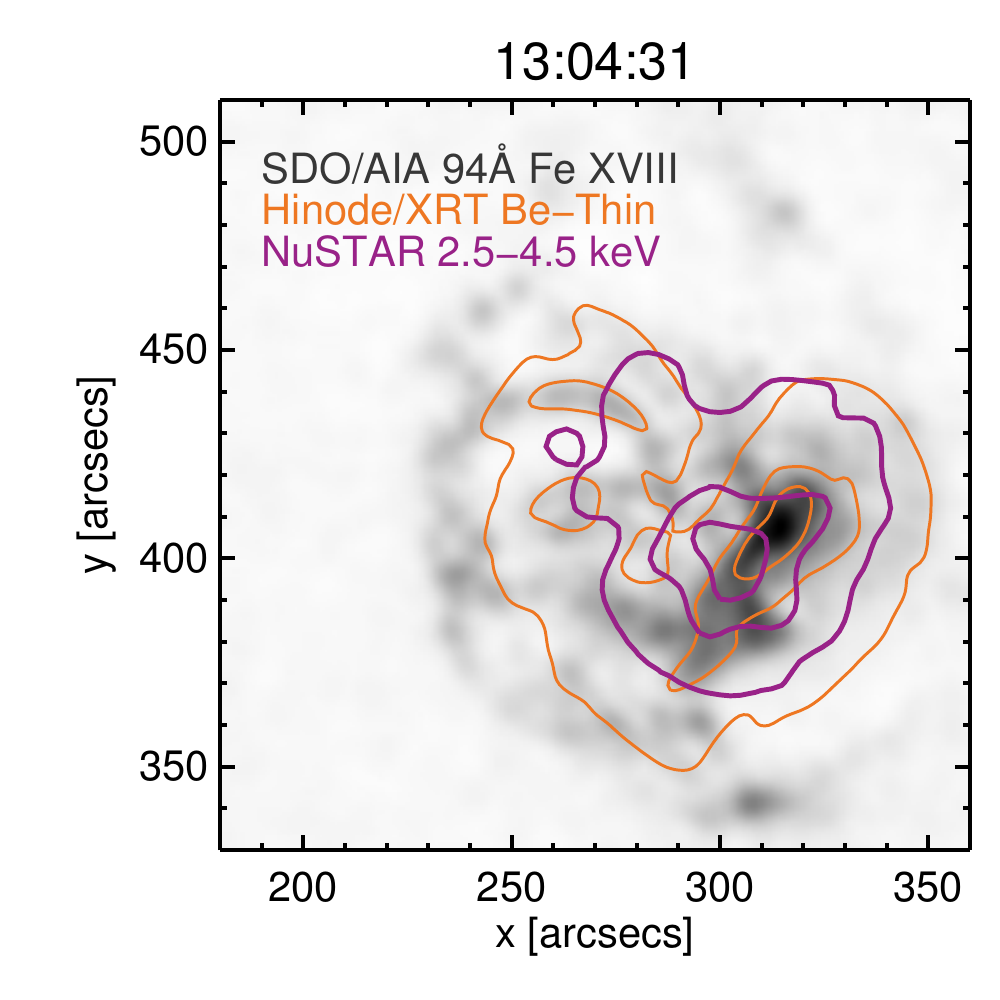}
	\caption{\label{fig:ns_shift}{\it SDO}/AIA 94\AA~\ion{Fe}{18} maps over plotted with shifted contours from {\it Hinode}/XRT (20\%, 50\%, 80\%; orange), and {\it NuSTAR} $2.5-4.5$ keV and $4.5-6.5$ keV emission (50\%, 70\%, 90\%; purple, turquoise). A constant offset correction was required for {\it Hinode}/XRT but a different one was determined for each {\it NuSTAR} pointing. 
For the two time intervals where {\it NuSTAR} only observed part of the AR (middle two panels), the alignment was done using the full map and to other features on the disk.}
\end{figure*}

Two of these pointings, the first and last, caught the whole AR but the other two only caught the lower part as they were observed at the edge of the detector, however this is the location of the heated loops during the microflares in Figure~\ref{fig:aiaxrtmap}. During some of the observations there was a change in the combination of Camera Head Units (CHUs) -- star trackers used to provide pointing information. In those such instances we used the time range that gave the longest continuous CHU combination, instead of the whole duration. Each required a different shift to match the {\it SDO}/AIA 94\AA~\ion{Fe}{18} map at that time, and all were within the expected 1\arcmin~offset \citep{2016BWG}. The alignment was straightforward for the {\it NuSTAR} maps which caught the whole region but was trickier for those with a partial observation. In those cases, second and third pointings, emission from another region (slightly to the south-west of AR12333) was used for the alignment. The resulting overlap of the aligned {\it Hinode}/XRT and {\it NuSTAR} images to {\it SDO}/AIA 94\AA~\ion{Fe}{18} are shown in Figure~\ref{fig:ns_shift}. The {\it NuSTAR} maps in Figure~\ref{fig:nsmap} reveal a similar pattern to the heating seen in EUV and SXR with {\it SDO}/AIA and {\it Hinode}/XRT: emission from the whole region before the microflare, with loops in the bottom right brightening as material is heated during the microflare, before fading as the material cools. 

\subsection{{\it NuSTAR} Spectral Fitting}

For each of the {\it NuSTAR} pointings we chose a region at the same location, and of the same area, as those used in the {\it SDO}/AIA and {\it Hinode}/XRT analysis, to produce spectra of the microflare heating. These are circular as the {\it NuSTAR} software can only calculate the response files for such regions, but do cover the flaring loop region (rectangular box, Figure~\ref{fig:aiaxrtmap}), and are shown in Figure~\ref{fig:nsmap}. The spectra and {\it NuSTAR} response files were obtained using the {\it NuSTAR} Data Analysis software v1.6.0. These were then fitted using the XSPEC \citep{1996ASPC..101...17A} software\footnote{\url{https://heasarc.gsfc.nasa.gov/xanadu/xspec/}}, which simultaneously fits the spectra from each telescope module (FPMA and FPMB) instead of just adding the data sets. We also use XSPEC as it allows us to find the best-fit solution using Cash statistics \citep{1979ApJ...228..939C} which helps with the non-Gaussian uncertainties we have for the few counts at higher temperatures.

\begin{figure*}[!t]
\centering
	\includegraphics[width=0.45\columnwidth]{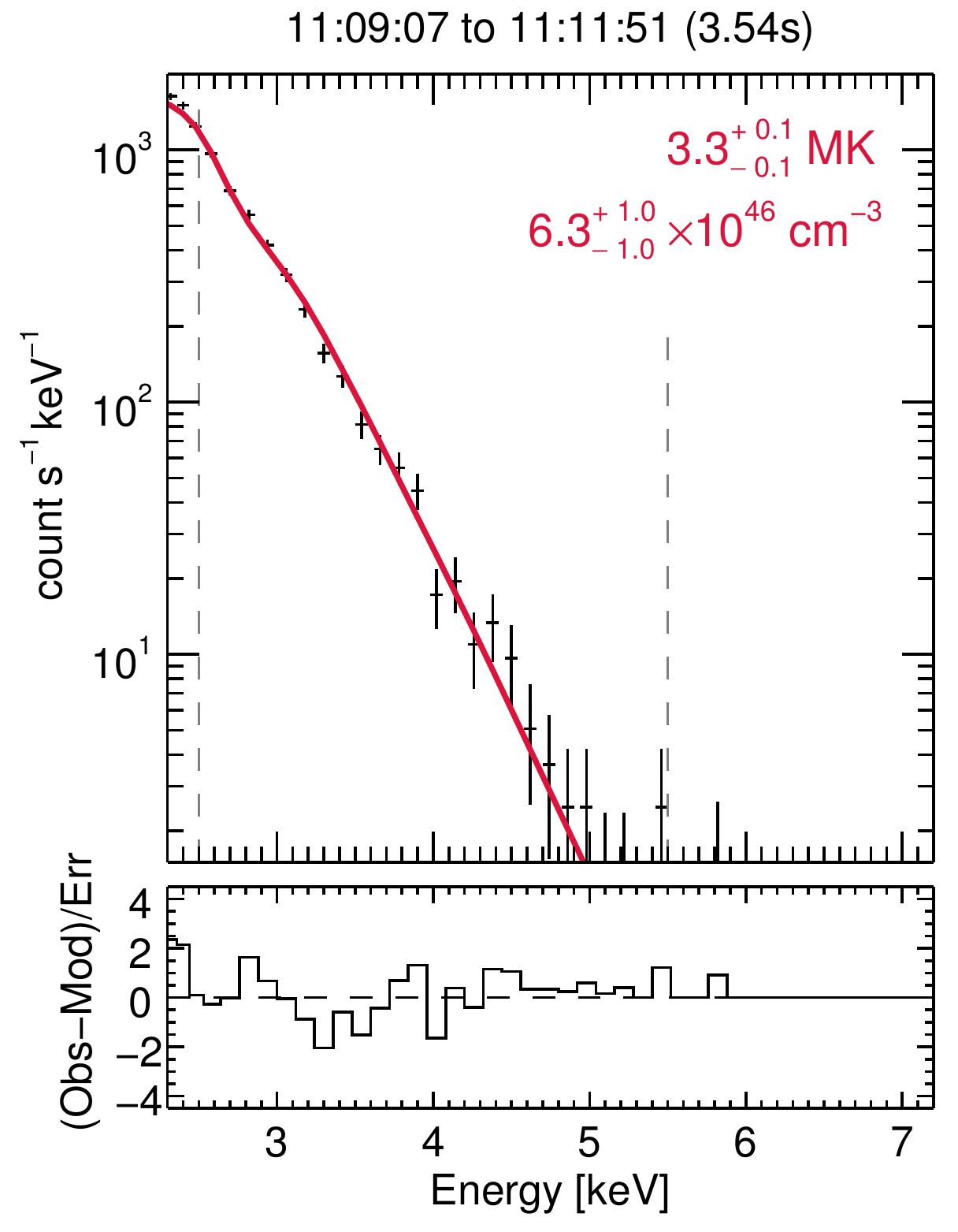}
 	\includegraphics[width=0.45\columnwidth]{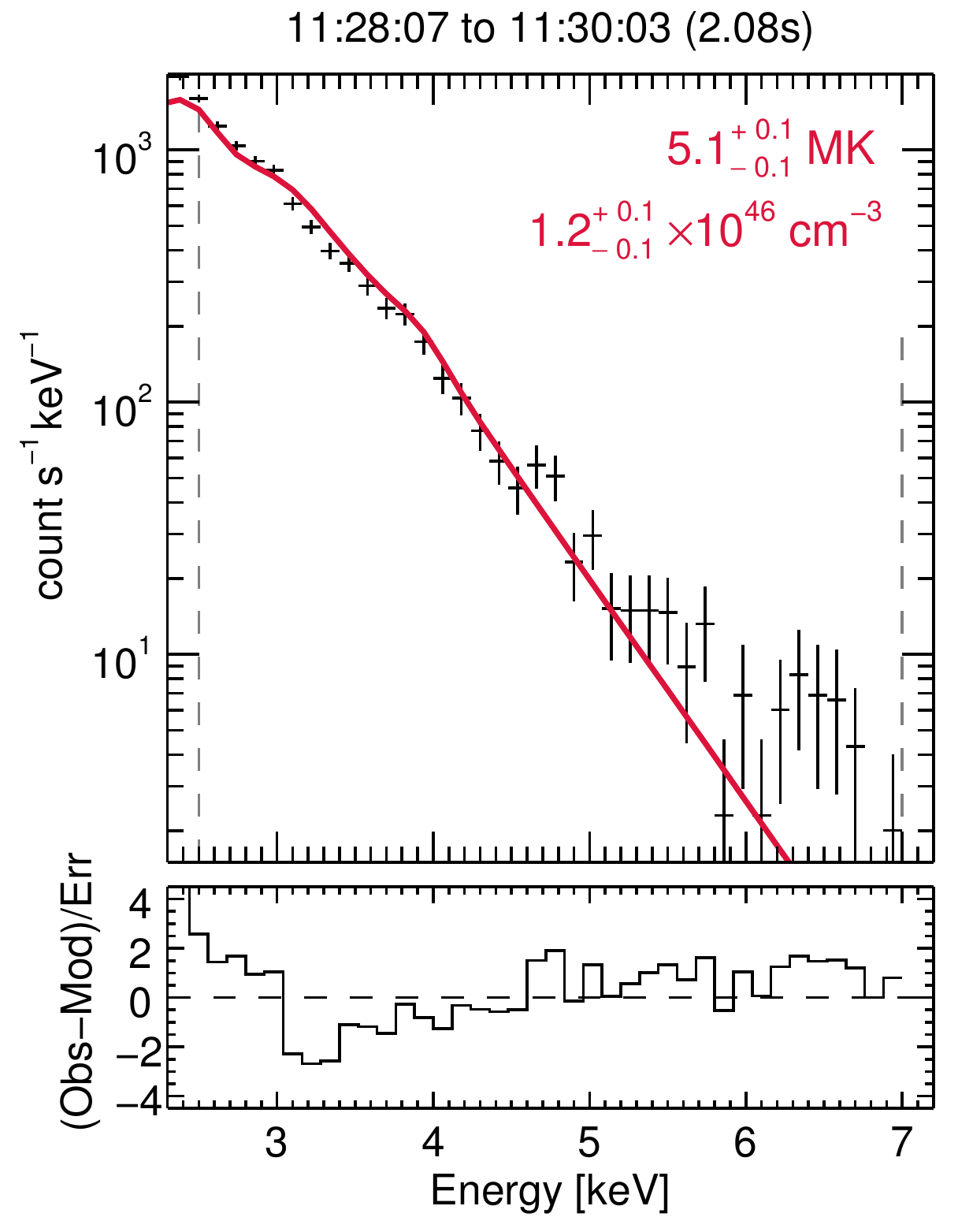}
	\includegraphics[width=0.45\columnwidth]{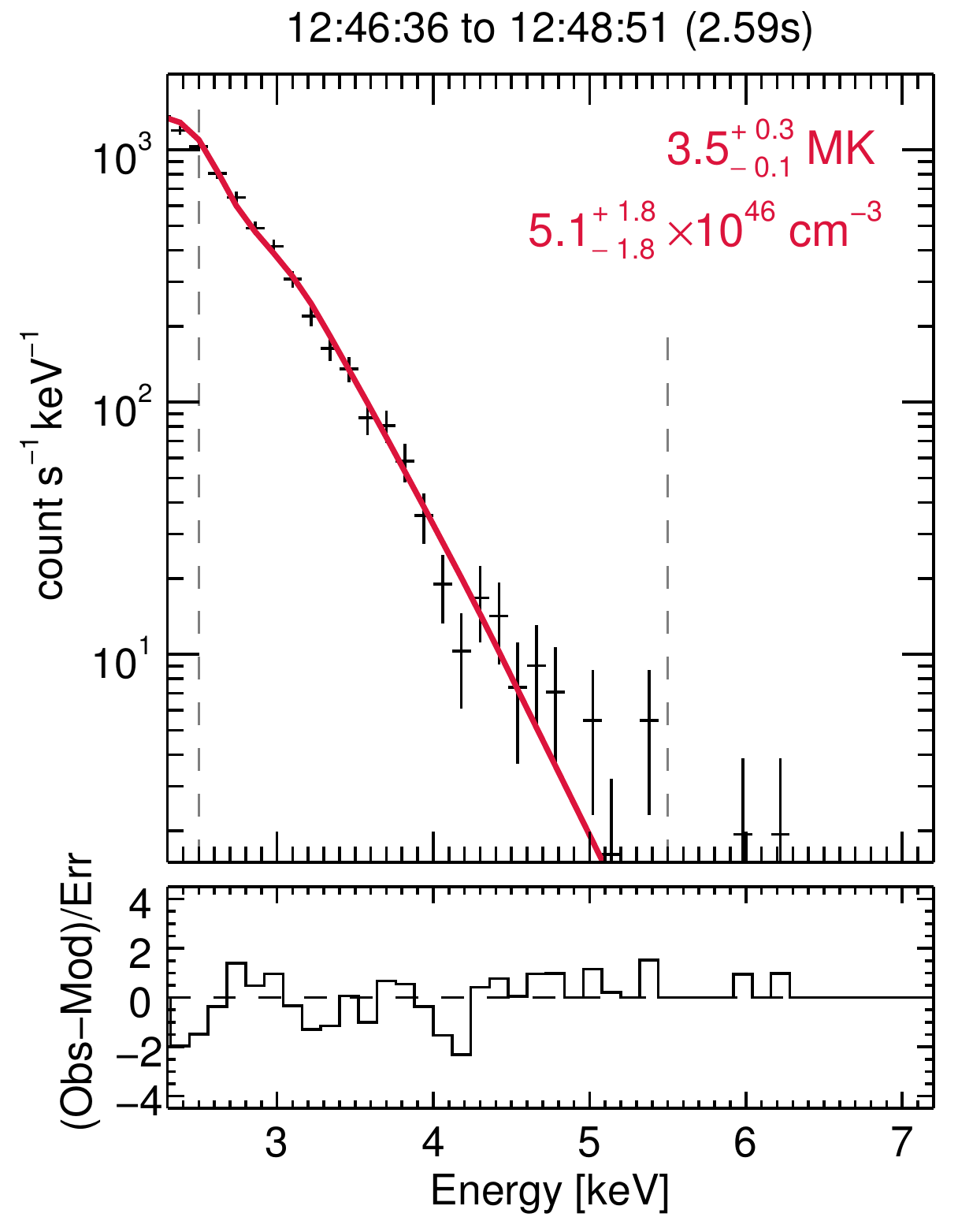}
	\includegraphics[width=0.45\columnwidth]{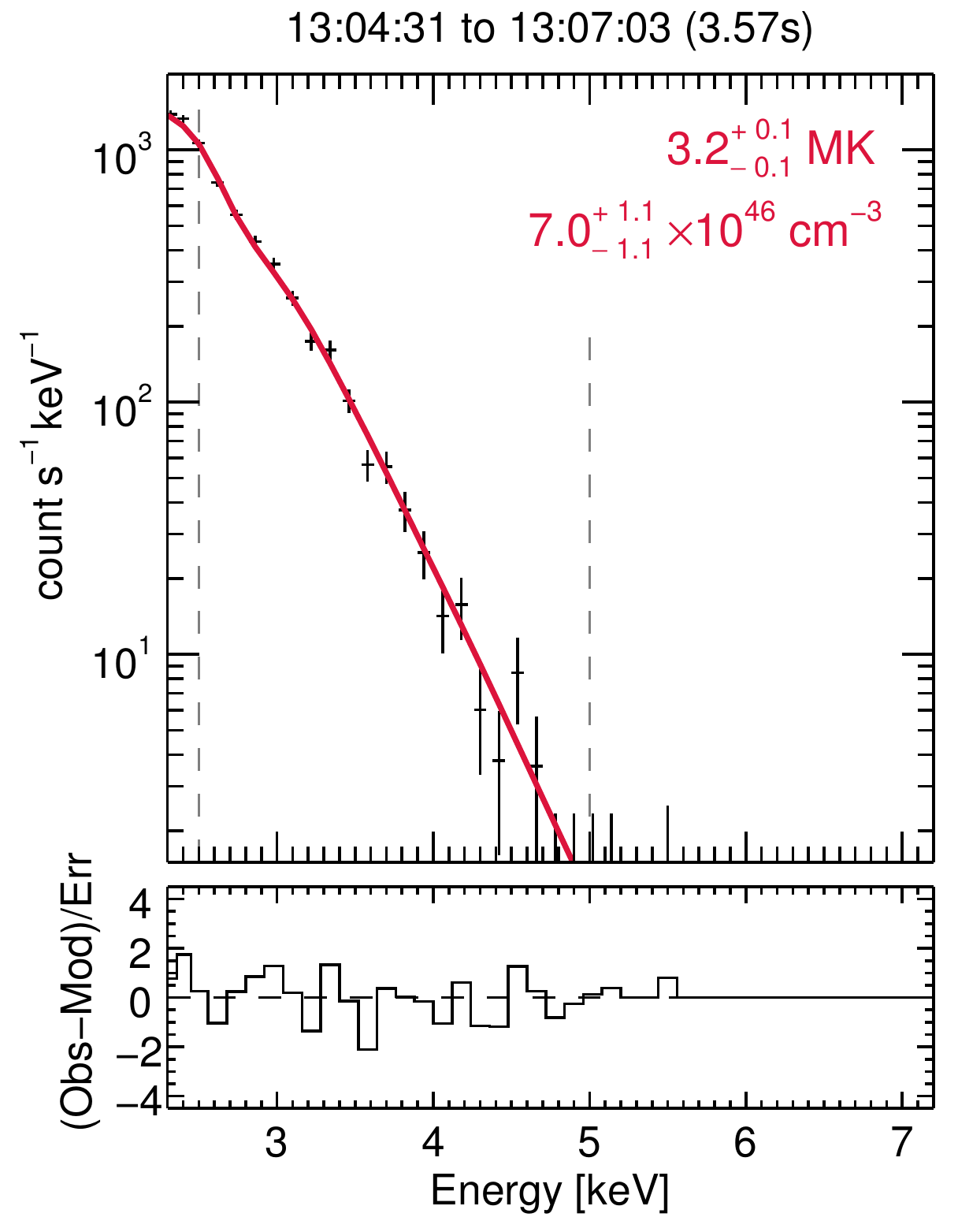}\\
	\caption{\label{fig:ns_xspec1}{\it NuSTAR} spectra for the regions shown in Figure~\ref{fig:ns_shift}, at different stages of flare evolution with time, increasing from left to right. The black data points show the combined data from FPMA and FPMB, and the red line shows the best fit thermal model. Note that the fit was performed to the data simultaneously and is only combined for plotting. The bottom panels show the residuals and the dashed vertical grey lines indicate the energy range over which the fit was performed (starting from the minimal usable energy of $2.5$ keV up where there are still substantial counts). The quoted uncertainties are with 90\% confidence.}
\end{figure*}

We fitted the spectra with a single thermal model, using the APEC model with solar coronal abundances \citep{1992ApJS...81..387F}, and the fit results are shown in Figure~\ref{fig:ns_xspec1}. For the first and fourth {\it NuSTAR} pointings, before and after the microflares, the spectra are well fitted by this single thermal model showing similar temperatures and emission measures ($3.3$ MK and $6.3\times10^{46}$ cm$^{-3}$, then $3.2$ MK and $7.0\times10^{46}$ cm$^{-3}$). Above $5$ keV there are very few counts and this is due to a combination of the low livetime of the observations (164s and 152s dwell time with about 2\% livetime fraction resulting in effective exposures of around 3.5s) and the high likelihood that the emission from this region peaked at this temperature before falling off very sharply at higher temperatures. These temperatures are similar to the quiescent ARs previously studied by {\it NuSTAR} \citep{2016ApJ...820L..14H}, although those regions were brighter and more numerous in the field of view, resulting in an order-of-magnitude worse livetime. The low livetime has the effect of limiting the spectral dynamic range, putting most of detected counts at the lower energy range, and no background or source counts at higher energies \citep{2016ApJ...820L..14H,2016BWG}.

\begin{figure*}
\centering
	\includegraphics[width=0.7\columnwidth]{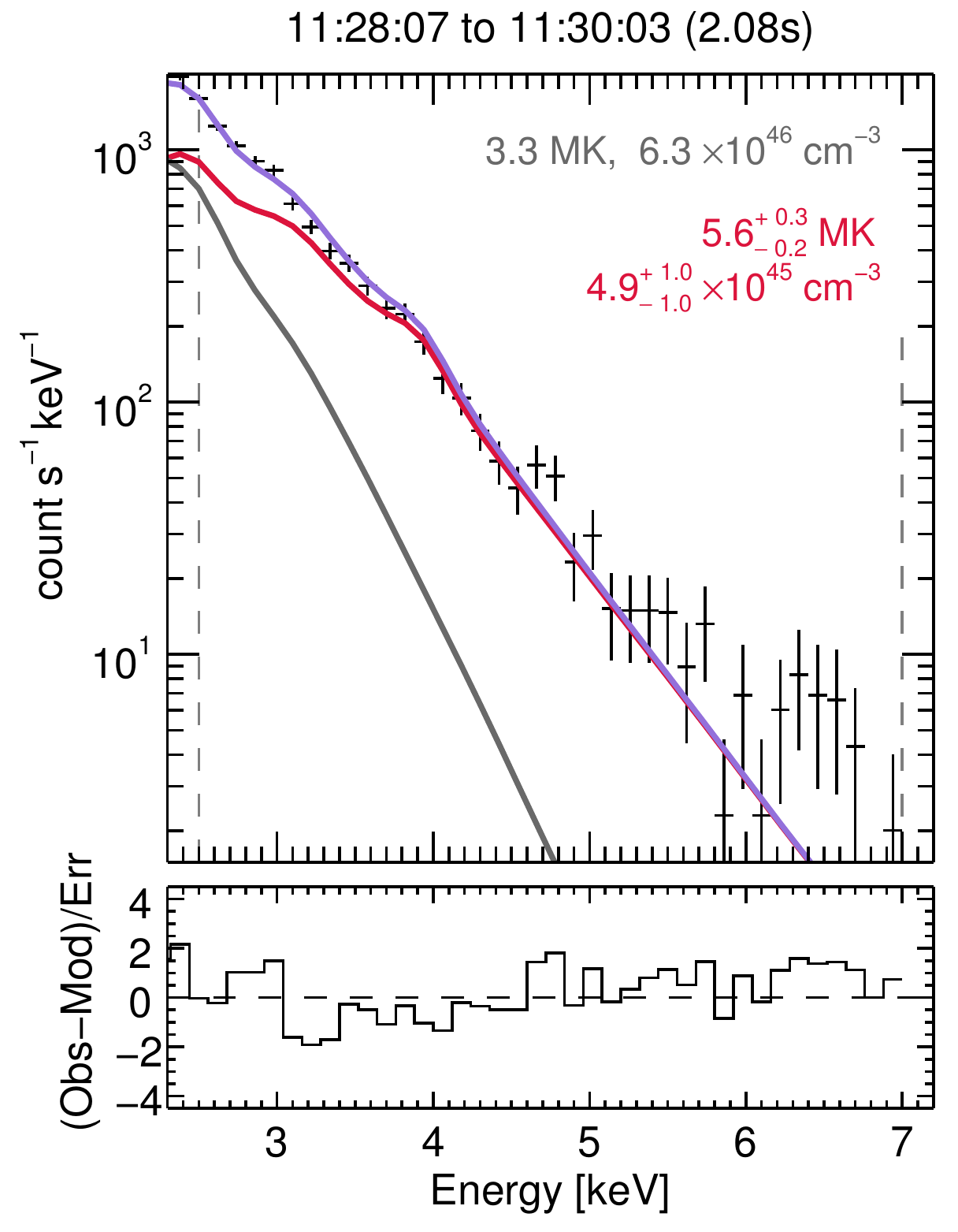}
	\includegraphics[width=0.7\columnwidth]{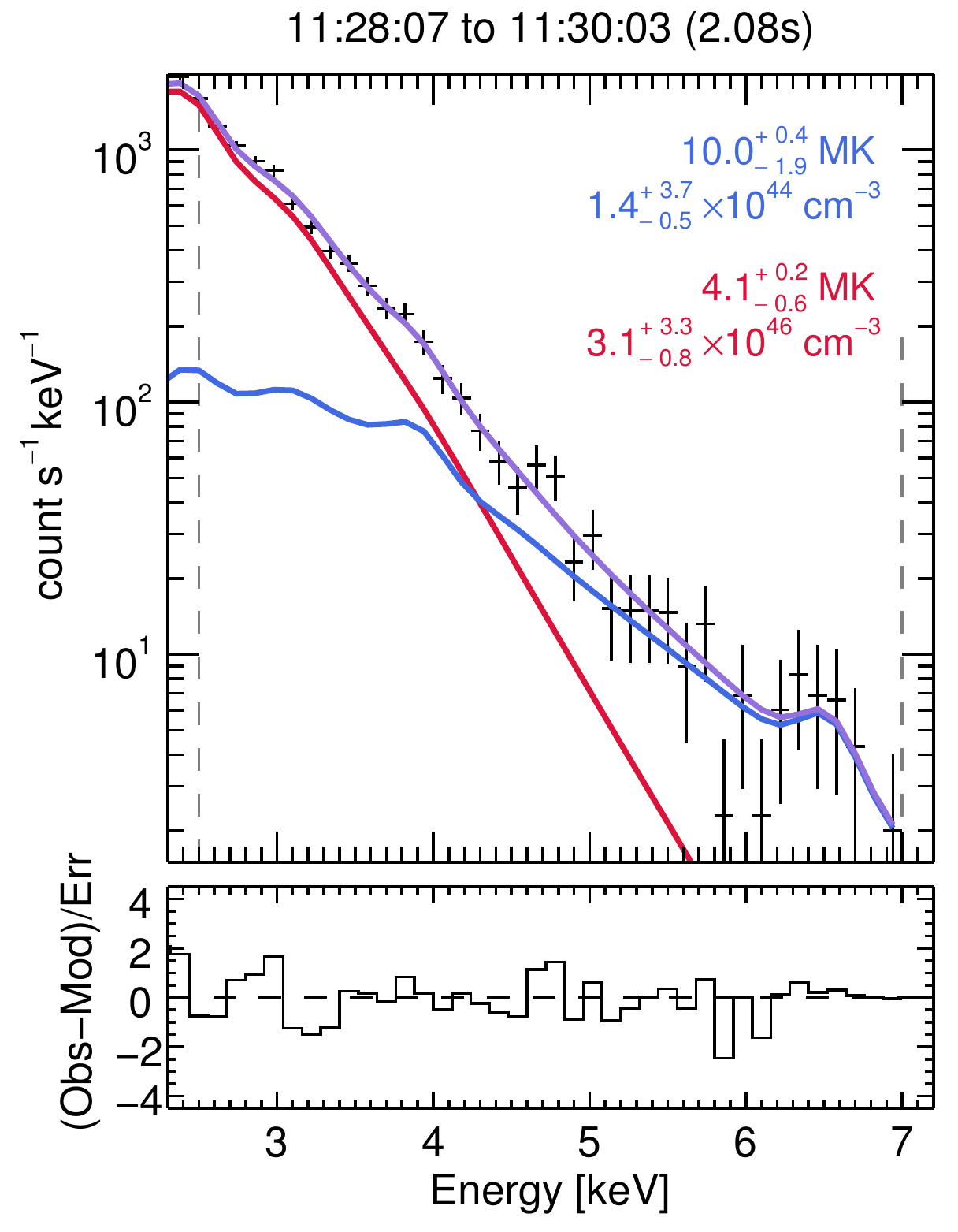}
	\caption{\label{fig:ns_xspec2}Additional model fits to the {\it NuSTAR} spectrum for the impulsive phase of the microflare. (Left) model of two thermals, one fixed using the parameters from the pre-flare observation (grey line), the second one (red) fitted. (Right) Model fitting two thermals. In both cases the total model is shown by the purple line and the black data points show the combined data from FPMA and FPMB. Note that the fit was performed to the data simultaneously and is only combined for plotting here. The quoted uncertainties are 90\% confidence levels.}
\end{figure*}

The two {\it NuSTAR} spectra from during the microflare, the second (impulsive phase) and third (decay phase, weaker peak) both show counts above $5$ keV and produce higher temperature fits ($5.1$ MK and $3.5$ MK). This is expected as there should be heating during the microflare, but neither fit matches the observed spectrum well, particularly during the impulsive phase. This shows that there is additional hot material during these times that a single-component thermal model cannot accurately characterise. For the spectrum during the impulsive phase, the second {\it NuSTAR} pointing, we tried adding additional thermal components to the fit, shown in Figure~\ref{fig:ns_xspec2}. We started by adding in a second thermal component fixed with the parameters from the pre-microflare spectrum, found from the first {\it NuSTAR} pointing (left spectrum in Figure~\ref{fig:ns_xspec1}), to represent the background emission. We did this as {\it NuSTAR}'s pointing changed during these two times (changing the part of the detector observing the region, and hence instrumental response) so we could not simply subtract the data from this pre-flare background time. The other thermal model component was allowed to vary and produced a slightly better fit to the higher energies and a higher temperature ($5.6$ MK). However this model still misses out counts at higher energies. 

So we tried another fit where the two thermal models were both allowed to vary and this is shown in the right of Figure~\ref{fig:ns_xspec2}. Here there is a substantially better fit to the data over the whole energy range, fitting a model of $4.1$ MK and $10.0$ MK. The hotter model does seem to match the bump in emission between $6$ and $7$ keV, which at these temperatures would be due to line emission from the Fe K-Shell transition \citep{2004ApJ...605..921P}. Although this model better matches the data, it produces substantial uncertainties, particularly in the emission measure. This is because it is fitting the few counts at higher energies which have a poor signal to noise. It should be noted that for the thermal model the temperature and emission measure are correlated and so the upper uncertainty on the temperature relates to the lower uncertainty on the emission measure, and vice versa. Therefore this uncertainty range covers a narrow diagonal region of parameter space, which we include later in Figure~\ref{fig:emd}. These fits do however seem to indicate that emission from material up to $10$ MK is present in this microflare and that the {\it NuSTAR} spectrum in this case is observing purely thermal emission. A non-thermal component could still be present, but the likely weak emission, combined with {\it NuSTAR}'s low livetime (limiting the spectral dynamic range), leaves this component hidden. Upper-limits to this possible non-thermal emission are calculated in \S\ref{sec:nontherm}.

From these spectral fits we estimated the {\it GOES} $1-8$~\AA~flux\footnote{\url{https://hesperia.gsfc.nasa.gov/ssw/gen/idl/synoptic/goes/goes_flux49.pro}} to be $5.3 \times 10^{-9}$ Wm$^{-2}$ for the impulsive phase, and $4.0 \times 10^{-9}$ Wm$^{-2}$ for the pre-flare time. This means that the background subtracted {\it GOES} class for the impulsive phase is equivalent to $\sim$A0.1 and would be slightly larger during the subsequent peak emission time.

\section{Multi-thermal Microflare Emission\label{sec:multit}}

The {\it NuSTAR} spectrum during the impulsive phase of the microflare clearly shows that there is a range of heated material, so to get a comprehensive view of this multi-thermal emission we recovered the differential emission measure (DEM) by combining the observations from {\it NuSTAR}, {\it Hinode}/XRT, and {\it SDO}/AIA. This is the first time these instruments have been used together to obtain a DEM.

\subsection{Comparison of {\it NuSTAR}, {\it Hinode}/XRT \& {\it SDO}/AIA}

\begin{figure*}
\centering
	\includegraphics[width=0.9\columnwidth]{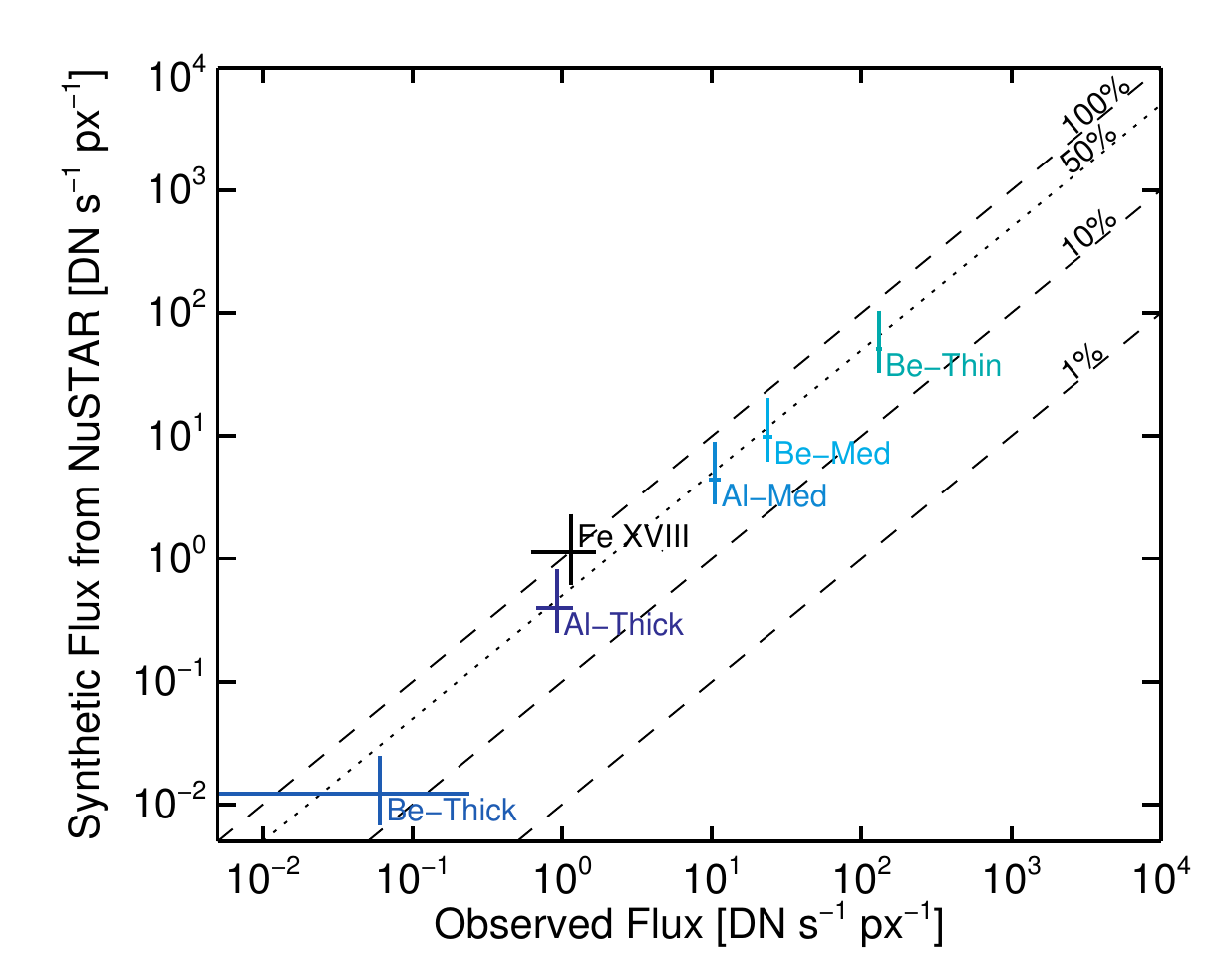}
	\includegraphics[width=0.9\columnwidth]{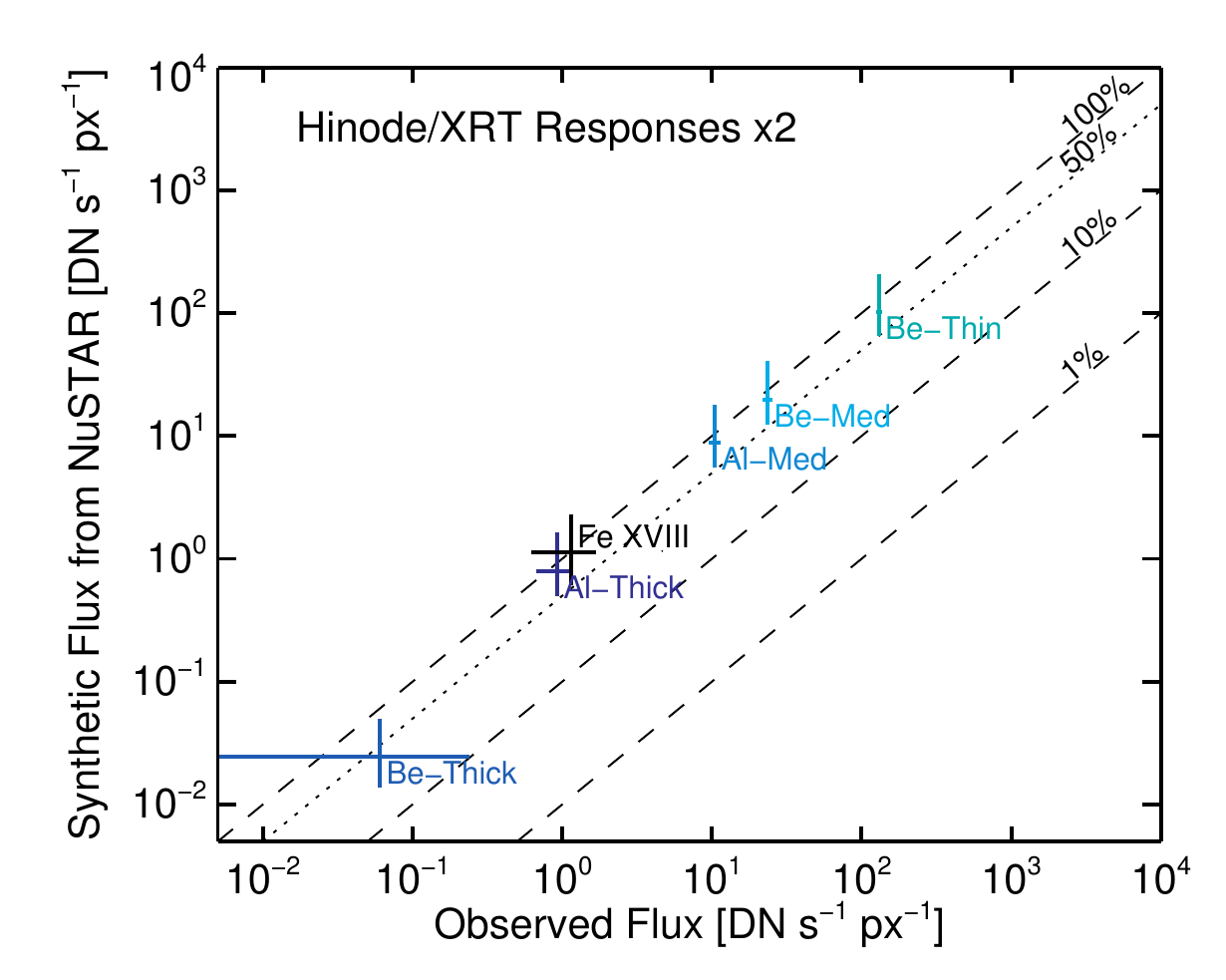}\\
	\includegraphics[width=0.9\columnwidth]{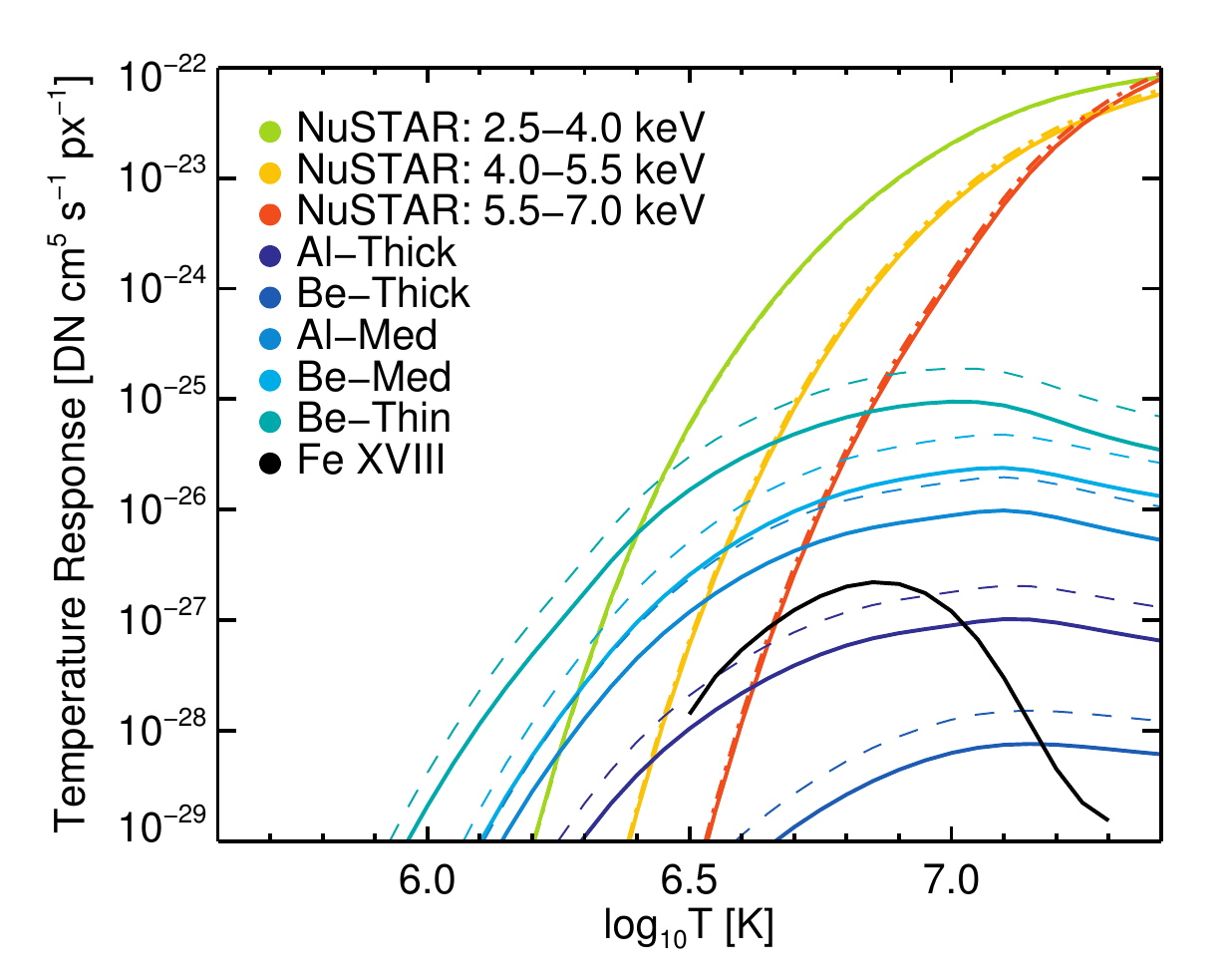}\\
		\caption{\label{fig:ns_syn}(Top) Comparison of {\it Hinode}/XRT and {\it SDO}/AIA 94\AA~\ion{Fe}{18} fluxes during the microflare's impulsive phase, to the synthetic values obtained from the {\it NuSTAR} spectral fit. (Bottom) The temperature response functions for {\it NuSTAR} (FPMA, solid; FPMB, dot-dash) for the region shown in Figure~\ref{fig:nsmap} (panel 2), {\it SDO}/AIA 94\AA~\ion{Fe}{18} (solid black), and {\it Hinode}/XRT (original, solid; $\times2$, dashed). This has been done using the standard {\it Hinode}/XRT responses (top left) and then multiplying them by a factor of two (top right) which gives values closer to the observed fluxes.}
\end{figure*}
 
To check the compatibility of the {\it NuSTAR}, {\it Hinode}/XRT, and {\it SDO}/AIA observations we compared the observed fluxes from {\it Hinode}/XRT, and {\it SDO}/AIA to synthetic fluxes obtained from the {\it NuSTAR} thermal fits. For the {\it NuSTAR} two thermal fit (Figure~\ref{fig:ns_xspec2}, right panel) we multiplied the emission measures by the {\it SDO}/AIA and {\it Hinode}/XRT temperature response functions at the corresponding temperatures, and then added the two fluxes together to get a value for each filter channel.

The {\it Hinode}/XRT temperature response functions were created using \texttt{xrt\_flux.pro} with a CHIANTI 7.1.3 \citep{1997A&AS..125..149D, 2013ApJ...763...86L} spectrum (\texttt{xrt\_flux713.pro}\footnote{\url{http://solar.physics.montana.edu/takeda/xrt_response/xrt_resp_ch713_newcal.html}}) with coronal abundances \citep{1992ApJS...81..387F}, and the latest filter calibrations that account for the time-dependent contamination layer present on the CCD \citep{2014SoPh..289.1029N}. The {\it SDO}/AIA temperature response functions are version six (v6; using CHIANTI 7.1.3) and obtained using \texttt{aia\_get\_response.pro} with the `chiantifix', `eve\_norm', and `timedepend\_date' flags. The comparison of the observed and synthetic fluxes are shown in Figure~\ref{fig:ns_syn}.

We found that the {\it SDO}/AIA 94\AA~\ion{Fe}{18} synthetic flux is near the observed value, as expected, however there is a consistent discrepancy for {\it Hinode}/XRT. The observed fluxes should match the synthetic fluxes from the {\it NuSTAR} spectral fits as they are sensitive to the same temperature range. Other authors have found similar discrepancies \citep{2011ApJ...728...30T, 2015ApJ...807..143C, 2015ApJ...806..232S} and there is the suggestion that the {\it Hinode}/XRT temperature response functions are too small by a factor of $\sim2 - 3$ \citep[see][]{2015ApJ...806..232S}. 
We have therefore multiplied the {\it Hinode}/XRT temperature response functions by a factor of two (Figure~\ref{fig:ns_syn}, top right) and find a closer match to the synthetic values derived from the {\it NuSTAR} spectral fits. The main effect of these larger temperature response functions is that it requires there to be weaker emission at higher temperatures to obtain the same {\it Hinode}/XRT flux.

\subsection{Differential Emission Measure}

Recovering the line-of-sight DEM, $\xi(T_{j})$, involves solving the ill-posed inverse problem, ${g}_{i}={\bf K}_{i,j}\xi(T)$, where $g_{i}$ [DN s$^{-1}$ px$^{-1}$] is the observable, and ${\bf K}_{i,j}$ is the the temperature response function for the $ith$ filter channel, and the $jth$ temperature bin. Numerous algorithms have been developed for the DEM reconstruction, and we use two methods to recover the DEM: Regularised Inversion\footnote{\url{https://github.com/ianan/demreg}} \citep[RI,][]{2012A&A...539A.146H}, and the \texttt{xrt\_dem\_iterative2.pro} method\footnote{\url{https://hesperia.gsfc.nasa.gov/ssw/hinode/xrt/idl/util/xrt_dem_iterative2.pro}} \citep[XIT,][]{2004ASPC..325..217G, 2004IAUS..223..321W}. 

The regularised inversion (RI) approach recovers the DEM by limiting the amplification of uncertainties using linear constraints. Uncertainties on the DEM are also found on both the DEM and temperature resolution (horizontal uncertainties), see \citet{2012A&A...539A.146H}. XIT is a forward-fitting iterative least-squares approach, using a spline model. Uncertainties in the final DEM are calculated with Monte Carlo (MC) iterations with input data perturbed by an amount randomly drawn from a Gaussian distribution with the standard deviation equal to the uncertainty in the observation. The resulting spread of these MC iterations indicates the goodness-of-fit. 

\begin{figure*}[!t]
\centering
	\includegraphics[width=2.0\columnwidth]{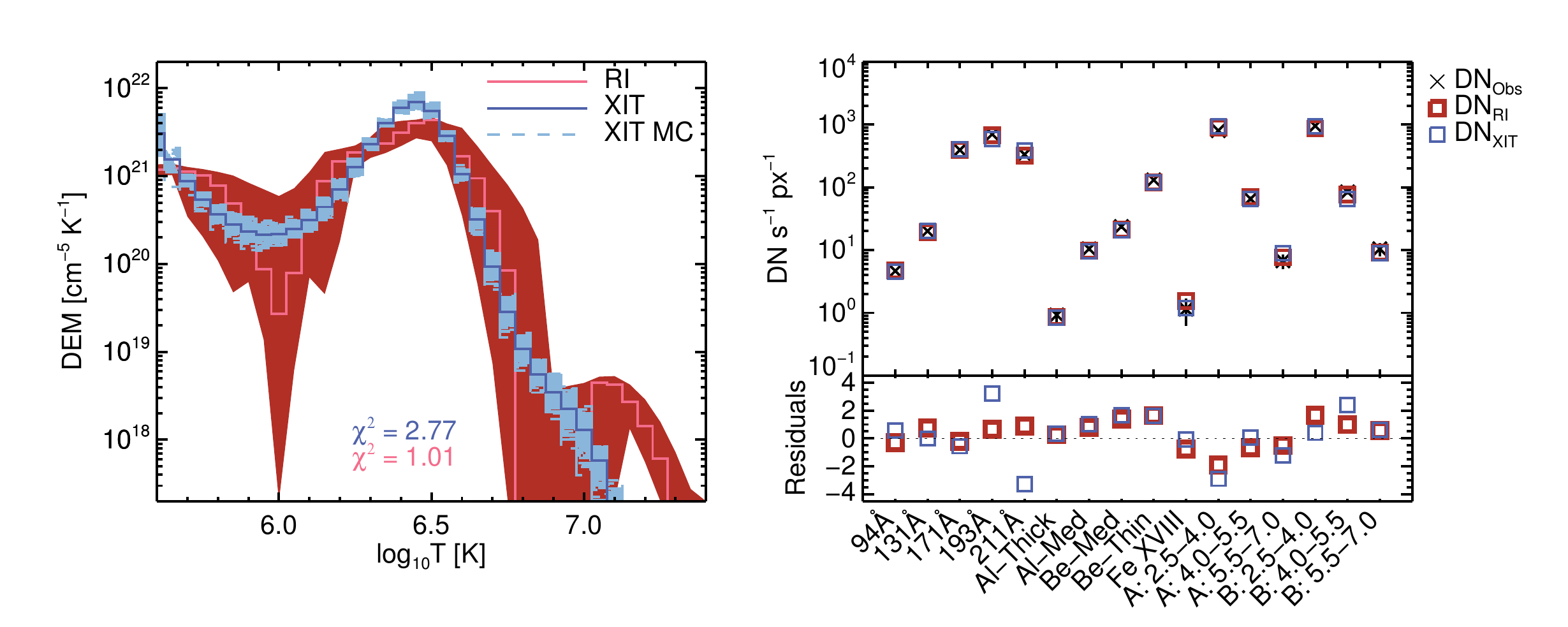}\\
	\includegraphics[width=2.0\columnwidth]{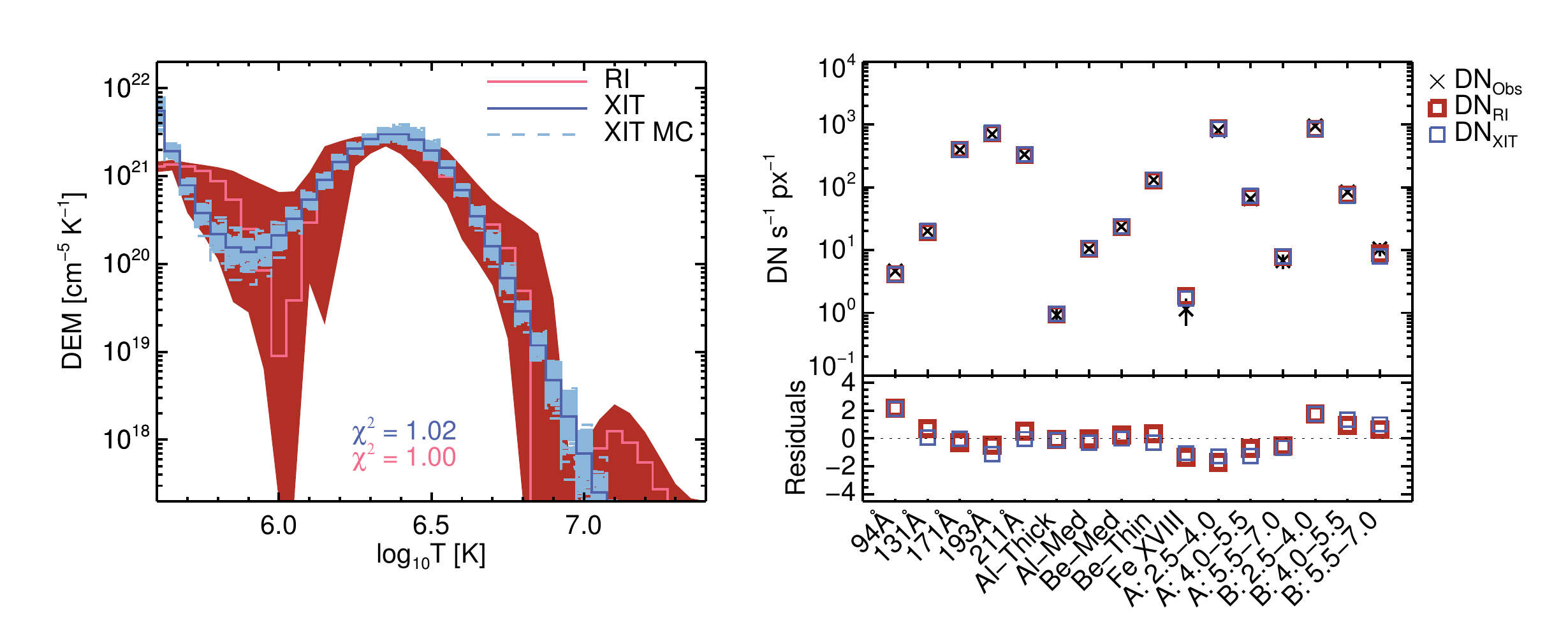}	
	\caption{\label{fig:dems}(Left) DEMs obtained during the impulsive phase of the microflare using {\it SDO}/AIA, {\it Hinode}/XRT and {\it NuSTAR} data. (Right) Residuals of the DEMs in data-space. The pink DEM (red error region) was obtained using the RI, and the blue (with 300 sky blue MC iterations) from XIT. The DEMs were calculated using both the standard {\it Hinode}/XRT temperature responses (top) as well as those multiplied by a factor of two (bottom).}
\end{figure*}

\begin{figure*}[!t]
\centering
	\includegraphics[width=2.0\columnwidth]{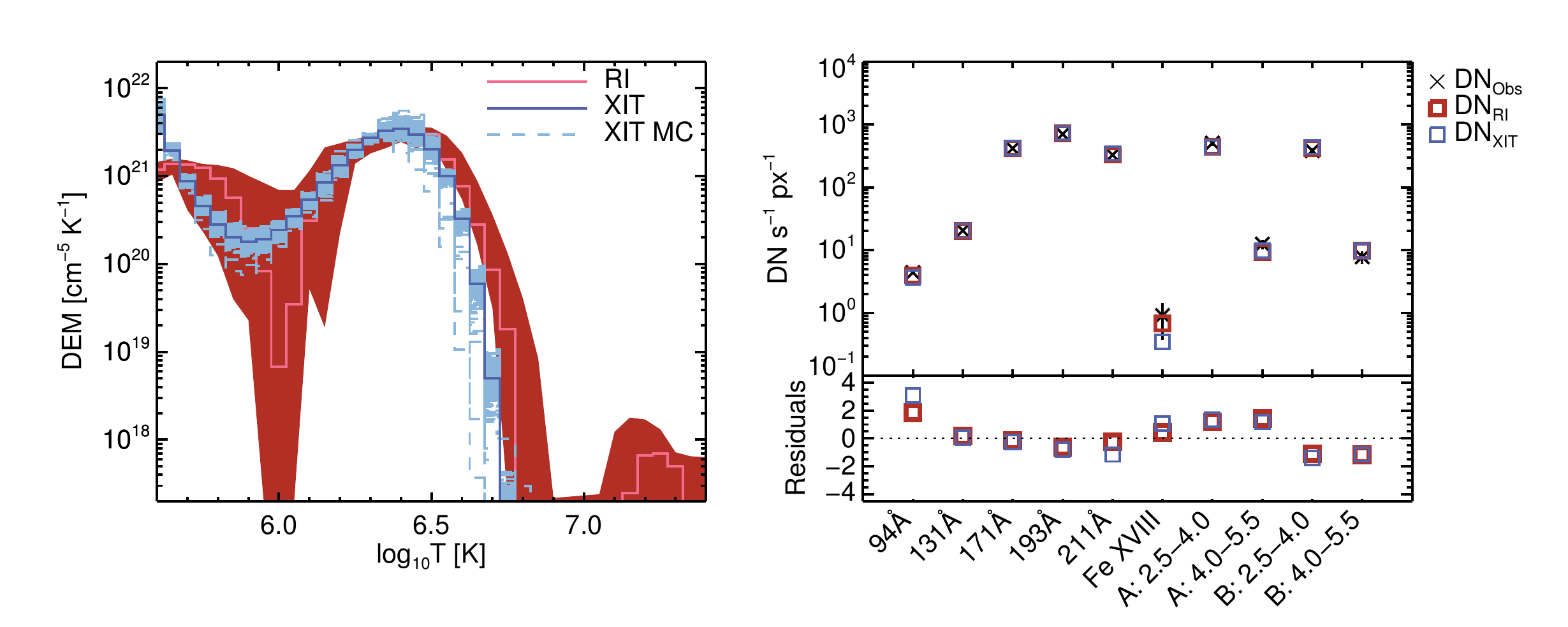}
	\includegraphics[width=2.0\columnwidth]{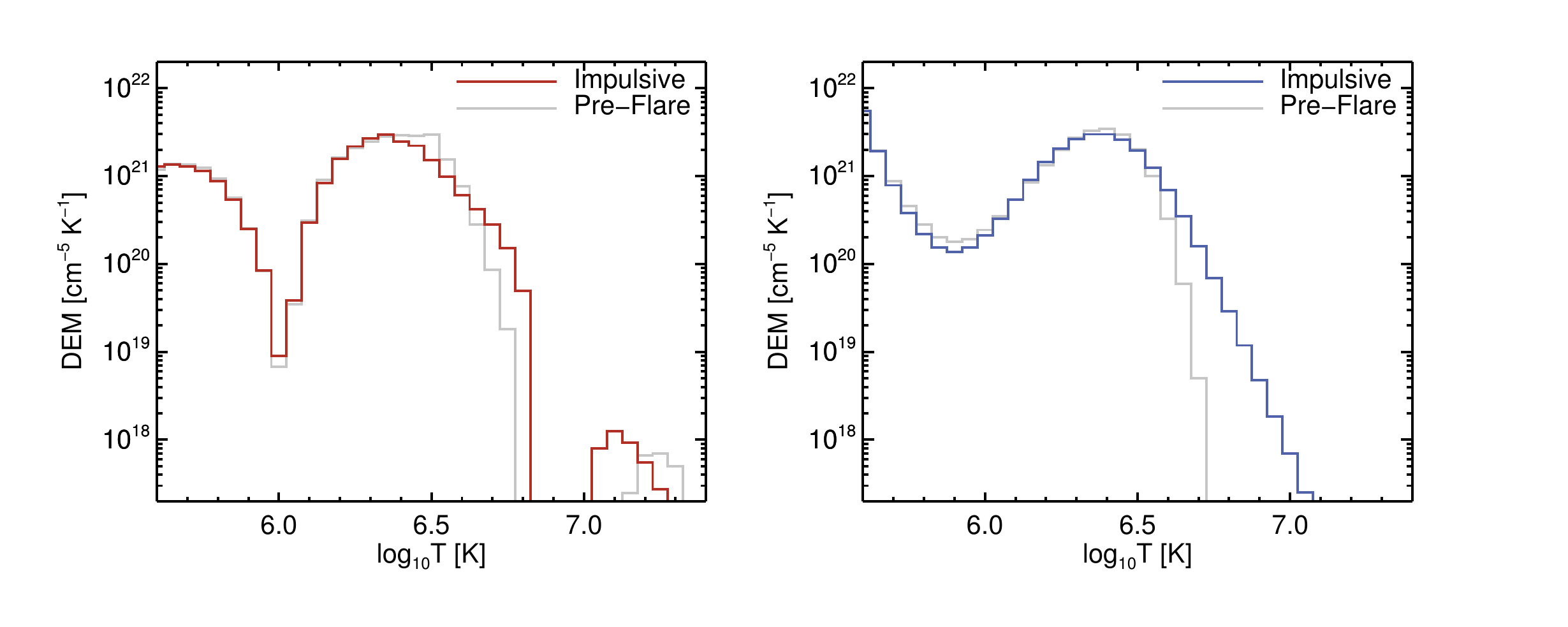}
	\caption{\label{fig:preflaredem}(Top Left) DEM obtained from the pre-flare phase ($\sim$ 11:10 UT) using {\it SDO}/AIA, and {\it NuSTAR} data. (Top Right) Residuals of the DEMs in data-space. (Bottom) The RI (left) and XIT (right) pre-flare DEMs shown in comparison to the impulsive-phase DEMs (Figure~\ref{fig:dems}, bottom row). The pre-flare DEMs peak at similar temperatures, and fall off more steeply than the impulsive-phase DEMs. The increase in the DEMs is due to the heating during the microflare.}
\end{figure*}

\begin{figure*}[!t]
	\centering
        	\includegraphics[width=2.0\columnwidth]{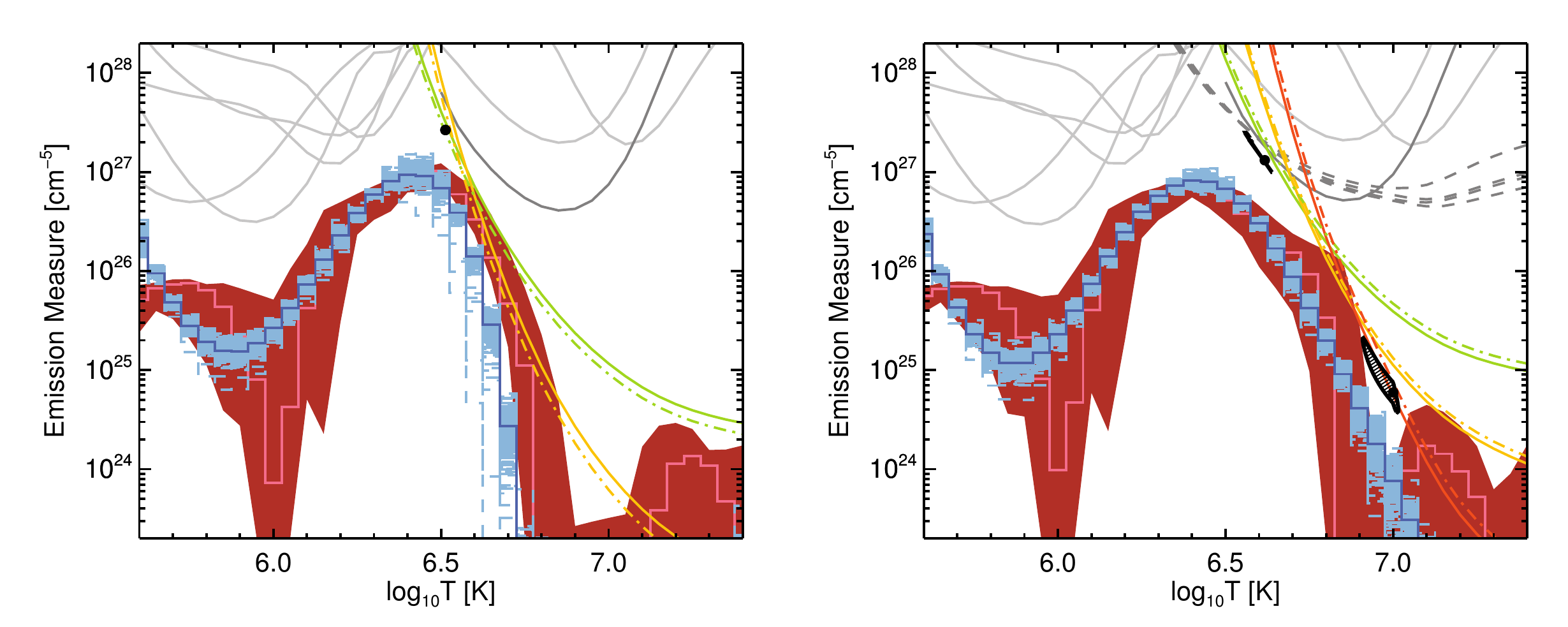}
	\caption{\label{fig:emd} Emission measure distribution obtained from the pre-flare (left) using {\it SDO}/AIA, and {\it NuSTAR} data, and the impulsive phase of the microflare (right) using {\it SDO}/AIA, {\it Hinode}/XRT and {\it NuSTAR} data with the {\it Hinode}/XRT responses multiplied by a factor of two. The EM loci curves for {\it NuSTAR} are shown in the same colours as in Figure~\ref{fig:ns_syn}; the {\it SDO}/AIA loci are plotted in grey, with 94\AA~\ion{Fe}{18} in dark grey; and {\it Hinode}/XRT loci are over plotted as dark grey dashed lines. The thermal fits from Figures~\ref{fig:ns_xspec1} and \ref{fig:ns_xspec2} are plotted as filled circles (black) with shaded 90\% confidence contours.}
\end{figure*}

For the DEM analysis we calculated the uncertainties on the {\it Hinode}/XRT and {\it SDO}/AIA data. The non-statistical photometric uncertainties for {\it Hinode}/XRT were calculated from \texttt{xrt\_prep.pro} \citep{2014SoPh..289.2781K}, and photon statistics calculated from \texttt{xrt\_cvfact.pro}\footnote{updated from CHIANTI 6.0.1 to CHIANTI 7.1.3 as part of this work} \citep{2011SoPh..269..169N, 2014SoPh..289.1029N}. The uncertainties on the {\it SDO}/AIA data were computed with \texttt{aia\_bp\_estimate\_error.pro} \citep{2012SoPh..275...41B}, and an additional 5\% systematic uncertainty was added in quadrature to both the {\it Hinode}/XRT and {\it SDO}/AIA data account for uncertainties in the temperature response functions. The {\it Hinode}/XRT and {\it SDO}/AIA data and uncertainties have been interpolated to a common time-step, and averaged over the {\it NuSTAR} observational duration. The uncertainty for the {\it NuSTAR} values in specific energy bands were determined as a combination of the photon shot noise and a systematic factor (of 5\%) to account for the cross-calibration between {\it NuSTAR}'s two telescope modules (FPMA and FPMB). The {\it NuSTAR} temperature response functions, for each energy range and telescope module (shown in Figure~\ref{fig:ns_syn}) were calculated using the instrumental response matrix for the regions shown in Figure~\ref{fig:nsmap}.

The resulting DEMs obtained for the impulsive phase are shown in Figure~\ref{fig:dems} (left) with the quality of the recovered DEM solution shown as residuals between the input, and recovered fluxes (right). XIT is used with the addition of $300$ MC iterations where outlier XIT MC solutions have been omitted. We have used all available filters with the exception of {\it Hinode}/XRT Be-Thick due to large uncertainties that are the result of a lack of counts (Figure~\ref{fig:timeprofile}), and {\it SDO}/AIA 335\AA~due to the observed long-term drop in sensitivity \citep[see Figure 1][]{2014SoPh..289.2377B}. The standard {\it Hinode}/XRT responses (Figure~\ref{fig:dems}, top) lead to disagreement between the two methods for DEM recovery, notably at the peak, and at higher temperatures ($\chi_{XIT}^{2} = 2.77$, $\chi_{RI}^{2} = 1.01$). Using the {\it Hinode}/XRT responses multiplied by a factor of two results in the methods having much better agreement ($\chi_{XIT}^{2} = 1.02$, $\chi_{RI}^{2} = 1.00$), and the DEM solutions result in smaller residuals, specifically in the {\it Hinode}/XRT filters. These final DEMs (Figure~\ref{fig:dems}, bottom) show a peak at $\sim3$ MK, and little material above $10$ MK.

To understand how much of this material has been heated out of the background during the microflare we performed DEM analysis for the pre-flare {\it NuSTAR} time ($\sim$ 11:10 UT). There is no {\it Hinode}/XRT data for this time so we determined the DEM using {\it NuSTAR} and {\it SDO}/AIA data. The DEMs for the pre-flare observations are shown in Figure~\ref{fig:preflaredem}. These DEMs for each method peak at a similar temperature ($\sim3$ MK), and fall off very sharply to $\sim5$ MK. During the microflare there is a clear addition of material up to $10$ MK (Figure~\ref{fig:preflaredem}, bottom).

We also represent the DEMs as the emission measure distributions (EMDs; $\xi(T)\text{d}T$) which allows us to compare the DEM results to the {\it NuSTAR} spectral fits, shown in Figure~\ref{fig:emd}. Here we have also over plotted the EM loci curves, $EM_{i} = g_{i} / K_{i}$, which are the upper-limits of emission based on an isothermal model, with the true solution lying below all of the EM loci curves. The {\it NuSTAR} thermal model fits are the isothermal (in the pre-flare phase) or two thermal (impulsive phase) fits to the multi-thermal plasma distribution, and so represent an approximation of the temperature distribution and emission measure. These models produce the expected higher emission measure values compared to the EMD, and are consistent with the EM loci curves.

\section{Microflare Energetics\label{sec:meng}}

\subsection{Thermal Energy\label{sec:thermal}}

For an isothermal plasma at a temperature T and emission measure EM, the thermal energy is calculated as

\begin{equation}
U_{T_{I}} = 3 k_{B}T \sqrt{EM f V} \quad\text{[erg]}
\label{eq:Uth}
\end{equation}

\noindent where $k_{B}$ is the Boltzmann constant, f, the filling factor, and V, the plasma volume \citep[e.g.][]{2008ApJ...677..704H}. Using the two thermal fit (Figure~\ref{fig:ns_xspec2}, right) we calculated the thermal energy during the impulsive phase, finding $U_{T_{I}} = 0.9 \times 10^{28}$ erg ($t_I = 116$ s). Here the equivalent loop volume, $V_E = fV$, was calculated as a volume of a cylinder enclosing only the flaring loop with length, L $\sim 50\arcsec$, and diameter, d $\sim6\arcsec$. This thermal energy includes both the microflare and background emission. We found the pre-flare energy (using fit parameters; Figure~\ref{fig:ns_xspec1}, left) as $U_{T_{I_{0}}} = 0.9 \times 10^{28}$~erg (and $t_{{I}_0} = 164$s). The resulting heating power during the microflare from the thermal fits to the {\it NuSTAR} spectra is then $P_{T_{I_{F}}} = U_{T_{I}}/t_{I} - U_{T_{I_{0}}}/t_{{I}_{0}} = 2.5 \times 10^{25}$ erg s$^{-1}$.

The thermal energy can also be estimated for a multi-thermal plasma using

\begin{equation}
U_{T} = 3 k_{B} V_E^{1/2} \frac{\int_{T}{T\xi_V(T)~dT}}{\sqrt{\int_{T}{\xi_V(T)~dT}}} \quad\text{[erg]}
\label{eq:Uth2}
\end{equation}

\noindent as described in \cite{2014ApJ...789..116I}, with the filling factor, $f = 1$, and $\xi_V(T)=n^2\mathrm{d}V/\mathrm{d}T$ in units of cm$^{-3}$ K$^{-1}$. For the RI and XIT DEM solutions we find values of $U_{T_{RI}} = 1.1 \times10^{28}$ erg, and $U_{T_{XIT}} = 1.2 \times10^{28}$ erg during the impulsive phase of the microflare. The pre-flare thermal energies we find $U_{T_{RI_0}} = 1.2 \times10^{28}$ erg, and $U_{T_{XIT_0}} = 1.2 \times10^{28}$ erg, and this then gives values of the heating power during the impulsive phase of the microflare as $P_{T_{RI_F}} = 2.3 \times10^{25}$ erg s$^{-1}$, and $P_{T_{XIT_F}} = 3.0 \times10^{25}$ erg s$^{-1}$. All of these approaches give a similar value for the heating, about $2.5 \times 10^{25}$ erg s$^{-1}$, over the microflare's impulsive period, and a summary of these values with uncertanties are given in Table~\ref{tab:AR}. It should be noted that these values are lower limits as the estimates neglect losses during heating.

\begin{deluxetable}{lcccc}[!h]
\tablecaption{Summary of thermal energies of AR12333.\label{tab:AR}}
\tablecolumns{4}
\tablewidth{0pt}
\tablehead{
\colhead{Method} &
\colhead{$U_{T_{0}}\tablenotemark{a}$} &
\colhead{$U_{T}\tablenotemark{b}$} &
\colhead{$P_{T_F}$} \\
\colhead{} &
\colhead{[$\times10^{28}$ erg]} &
\colhead{[$\times10^{28}$ erg]} &
\colhead{[$\times10^{25}$ erg s$^{-1}$]} 
}
\startdata
{\it NuSTAR} fit & $0.9^{+ 0.1}_{- 0.1}$ & $0.9^{+ 0.6}_{- 0.2}$ & $2.5^{+ 5.4}_{- 1.6}$ & \\
RI & $1.2^{+ 0.1}_{- 0.1}$ & $1.1^{+ 0.1}_{- 0.1}$  & $2.3^{+ 0.9}_{- 1.0}$ &  \\
XIT & $1.2^{+ 0.1}_{- 0.1}$ & $1.2^{+ 0.1}_{- 0.1}$  & $3.0^{+ 0.6}_{- 0.7}$ \\
\enddata
\tablecomments{The uncertainties on the energies and power derived from the {\it NuSTAR} fit are $2.7\sigma$ (90\% confidence), and those from RI/XIT are $1\sigma$.}
\tablenotetext{a}{$164$s observation}
\tablenotetext{b}{$116$s observation}
\end{deluxetable}

From the analysis of $25,705$ {\it RHESSI} events, \citep[Table 1][]{2008ApJ...677..704H} microflare thermal energies were found to range from $U_T = 10^{26 - 30}$ erg ($5\% - 95\%$ range; from a $16$s observation). This is equivalent to $P_{T} = 6.3\times10^{24-28}$ erg s$^{-1}$, and therefore the thermal power from our {\it NuSTAR} microflare is in the lower range of {\it RHESSI} observations. This is as expected as {\it NuSTAR} should be able to observe well beyond {\it RHESSI}'s sensitivity limit to small microflares.

\begin{figure*}[!t]
	\centering
         \includegraphics[width=2.0\columnwidth]{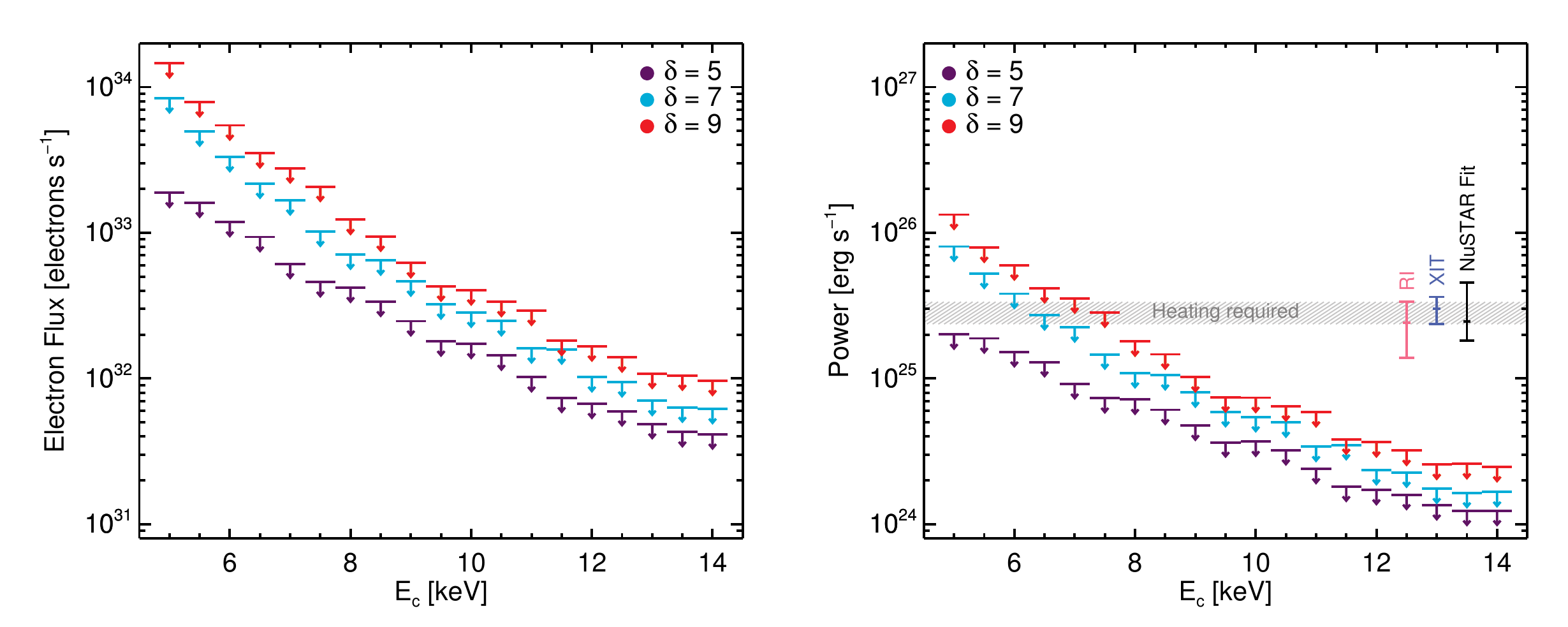}
         \caption{\label{fig:nontherm}Non-thermal upper-limits as a function of $E_c$ and $\delta$ plotted in terms of non-thermal electron flux, $N_{N}$ (left), and non-thermal power, $P_{N}$ (right). The three estimates for the thermal power: $P_{T_{I_{F}}}$, black; $P_{T_{RI_F}}$, pink; and $P_{T_{XIT_F}}$, blue, are plotted with $1\sigma$ uncertainties. The grey shaded region represents the required heating power, consistent with all three estimates.}
\end{figure*}

\subsection{{\it NuSTAR} Non-thermal Limits\label{sec:nontherm}}

As the {\it NuSTAR} spectrum in Figure~\ref{fig:ns_xspec2} is well fitted by a purely thermal model we can therefore find the upper-limits of the possible non-thermal emission. This approach allows us to determine whether the accelerated electrons could power the observed heating in this microflare. We used the thick-target model of a power-law electron distribution above a low-energy cut-off E$_{c}$ [keV] given by

\begin{equation}
F(E>E_{c}) \propto E^{-\delta}
\end{equation}

\noindent where $\delta$ is the power-law index, and the power in this non-thermal distribution is given by

\begin{equation}
    P_{N}(>E_{c}) = 1.6\times10^{-9} \frac{\delta - 1}{\delta - 2} N_{N} E_{c} \quad \text{[erg s}^{-1}\text{]}
\end{equation}

\noindent where $N_N$ is the non-thermal electron flux [electrons s$^{-1}$].

We determined the upper-limits on $N_{N}$ (and $P_{N}$) for a set of $\delta$ ($\delta = 5,7,9$), and $E_{c}$ consistent with a null detection in the {\it NuSTAR} spectrum. We performed this by iteratively reducing the model electron flux $N_{N}$ until there were less than 4 counts $>7$ keV, consistent with a null detection to $2\sigma$ \citep{1986ApJ...303..336G}. We also ensured the number of counts $\le 7$ keV are within the counting statistics of the observed counts. For each iteration we generated the X-ray spectrum for the two-component fitted thermal model (Figure~\ref{fig:ns_xspec2}, right) and added to this the non-thermal X-ray spectrum for our chosen $\delta$, $E_{c}$, and $N_{N}$, calculated using \texttt{f\_thick2.pro}\footnote{\url{https://hesperia.gsfc.nasa.gov/ssw/packages/xray/idl/f_thick2.pro}} \citep[see][]{2011SSRv..159..107H}. This was then folded through the {\it NuSTAR} response to generate a synthetic spectrum \citep[as discussed in][]{2016ApJ...820L..14H}. The upper-limits are shown in Figure~\ref{fig:nontherm} along with the three estimates of the thermal power for the background-subtracted flare, $P_{T_{I_F}}$ (``NuSTAR Fit'', black), $P_{T_{RI_F}}$ (pink), and $P_{T_{XIT_F}}$ (blue). For a flatter spectrum of $\delta = 5$ barely any of the upper-limits are consistent with the required heating power. With a steeper spectrum, $\delta \ge 7$, a cut-off $E_{c} \lesssim 7$ keV is consistent with the heating requirement. These steep spectra indicate that the bulk of the non-thermal emission would need to be at energies close to the low energy cut-off to be consistent with the observed {\it NuSTAR} spectrum. If we instead consider some of the counts in the $6 - 7$ keV range to be non-thermal (e.g. the excess above thermal model in the left panel in Figure~\ref{fig:ns_xspec2}) then we would obtain higher non-thermal power, about a factor of $0.5$ larger. However this would only substantially effect the steep non-thermal spectra ($\delta \ge 7$) as flatter models would be inconsistent with the data below $7$ keV.

We can again compare the microflare studied here to non-thermal energetics derived from {\it RHESSI} microflare statistics. \citet{2008ApJ...677..704H} report non-thermal parameters of $\delta = 4 - 10$, $E_{c} = 9 - 16$ keV, and the non-thermal power ranges from $P_{N}(\ge E_{c}) = 10^{25 - 28}$ erg s$^{-1}$. The largest upper-limits {\it NuSTAR} produces for this microflare are again at the the edge of {\it RHESSI}'s sensitivity. In a previous study of nanoflare heating \citet{2014Sci...346B.315T} investigated the evolution of chromospheric and transition region plasma from {\it IRIS} observations using RADYN nanoflare simulations. This is one of the few non-thermal nanoflare studies and they reported that heating occurred on time-scales of $\lesssim30$s characterized by a total energy $\lesssim10^{25}$ erg, and $E_{c} \sim 10$ keV. The simulated electron beam parameters in this {\it IRIS} event are consistent with the {\it NuSTAR} derived parameters, but in a range insufficient to power the heating in our microflare.

\section{Discussion and Conclusions}

In this paper we have presented the first joint observations of a microflaring AR with {\it NuSTAR}, {\it Hinode}/XRT, and {\it SDO}/AIA. During the impulsive start, the {\it NuSTAR} spectrum shows emission up to $10$ MK, indicating that even in this $\sim$A0.1 microflare substantial heating can occur. This high temperature emission is confirmed when we recover DEMs using the {\it NuSTAR}, {\it Hinode}/XRT, and {\it SDO}/AIA data. These instruments crucially overlap in temperature sensitivity, with {\it NuSTAR} able to constrain and characterise the high temperature emission which is often difficult for other instruments to do alone.

In this event we find the {\it Hinode}/XRT temperature response functions are a factor of two too small, suggesting that it would normaly overestimate the contribution from high temperature plasma in this microflare.

Overall we find the instantaneous thermal energy during the microflare to be $\sim10^{28}$ erg, once the pre-flare has been subtracted this equates to a heating rate of $\sim2.5 \times10^{25}$ erg s$^{-1}$ during the impulsive phase of this microflare. This is comparable to some of the smallest events observed with {\it RHESSI}, although {\it RHESSI} did not see this microflare as its indirect imaging was dominated by the brighter ARs elsewhere on the disk.

Although no non-thermal emission was detected, we can place upper-limits on the possible non-thermal component. We find that we would need a steep ($\delta \ge 7$) power-law down to at least $7$ keV to be able to power the heating in this microflare. This is still consistent with this small microflare being physically similar to large microflares and flares, but this would only be confirmed if {\it NuSTAR} detected non-thermal emission. To achieve this, future {\it NuSTAR} observations need to be made with a higher effective exposure time. For impulsive flares this cannot be achieved with longer duration observations, but only with higher livetimes. Observing the Sun when there are weaker or fewer ARs on the disk would easily achieve this livetime increase, conditions that have occurred since this observation and will continue through solar minimum.

These observations would greatly benefit from new, more sensitive, solar X-ray telescopes such as the {\it FOXSI} \citep{2014ApJ...793L..32K} and {\it MaGIXS} \citep{2011SPIE.8147E..1MK} sounding rockets, as well as the {\it MinXSS} CubeSats \citep{2016JSpRo..53..328M}. New data combined with {\it NuSTAR} observations during quieter periods of solar activity should provide detection of the high-temperature and possible non-thermal emission in even smaller microflares which should, in turn, provide a robust measure of their contribution to heating coronal loops in ARs.

\acknowledgments

This paper made use of data from the {\it NuSTAR} mission, a project led by the California Institute of Technology, managed by the Jet Propulsion Laboratory, funded by the National Aeronautics and Space Administration. We thank the {\it NuSTAR} Operations, Software and Calibration teams for support with the execution and analysis of these observations. This research made use of the {\it NuSTAR} Data Analysis Software (NuSTARDAS) jointly developed by the ASI Science Data Center (ASDC, Italy), and the California Institute of Technology (USA). {\it Hinode} is a Japanese mission developed and launched by ISAS/JAXA, with NAOJ as domestic partner and NASA and STFC (UK) as international partners. It is operated by these agencies in co-operation with ESA and the NSC (Norway). The Atmospheric Imaging Assembly on the {\it Solar Dynamics Observatory} is part of NASA’s Living with a Star program. CHIANTI is a collaborative project involving George Mason University, the University of Michigan (USA) and the University of Cambridge (UK). This research made extensive use of the IDL Astronomy Library, the SolarSoft IDL distribution (SSW), and NASA's Astrophysics Data System. 

PJW was supported by an EPSRC/Royal Society Fellowship Engagement Award (EP/M00371X/1) and IGH was supported by a Royal Society University Fellowship. MK and SK were supported by the Swiss National Science Foundation (project number 200021-140308 and 200020-169046). AJM was supported by NASA Earth and Space Science Fellowship award NNX13AM41H. This work was also supported by NASA grants NNX12AJ36G, NNX14AG07G. 

The authors thank the International Space Science Institute (ISSI) for support for the team ``New Diagnostics of Particle Acceleration in Solar Coronal Nanoflares from Chromospheric Observations and Modeling'', where this work benefited from productive discussions. The authors also thank P. J. A. Sim{\~o}es, S. H. Saar, K. K. Reeves, and J. K. Vogel for their valuable comments.
\facilities{NuSTAR, Hinode (XRT), SDO (AIA), GOES}

\end{document}